\documentclass[12pt]{article}
\usepackage{epsfig}
\usepackage{axodraw}
  
\newcommand{\mrm}[1]{\mathrm{#1}}
\newcommand{\as}{\alpha_{\mrm{s}}}
\newcommand{\qbar}{\mrm{\overline{q}}}
\newcommand{\pbar}{\mrm{\overline{p}}}
\newcommand{\cbar}{\mrm{\overline{c}}}
\newcommand{\ubar}{\mrm{\overline{u}}}
\newcommand{\pt}{p_{\perp}}
\newcommand{\kt}{k_{\perp}}
\newcommand{\shat}{\hat{s}}
\newcommand{\that}{\hat{t}}
\newcommand{\uhat}{\hat{u}}
\newcommand{\py}{\textsc{Pythia}}
\newcommand{\lqcd}{\Lambda_{\mrm{QCD}}}
\newcommand{\q}{Q^2}
\newcommand{\m}{m^2}
\newcommand{\mz}{m_{\mrm{Z}}^2}
\newcommand{\pplus}{p_{+}}
\newcommand{\pminus}{p_{-}}
\newcommand{\pz}{p_{z}}
\newcommand{\eplus}{e^{+}}
\newcommand{\eminus}{e^{-}}
\newcommand{\dint}{\mrm{d}}
\newcommand{\PS}{\mrm{PS}}

\newlength{\abstwidth}
\begin{document}

\sloppy

\pagestyle{empty}
 
\begin{flushright}
hep-ph/0401121\\
LU TP 04 - 01 \\
January , 2004
\end{flushright}
 
\vspace{\fill}
 
\begin{center}
{\LARGE\bf Perturbative and Nonperturbative}\\[3mm]
{\LARGE\bf Effects in Transverse Momentum}\\[4mm]
{\LARGE\bf Generation}\\[10mm]
{\large  Master of Science Thesis by Erik Thom\'{e}}\\ [2mm]
{\large  Thesis advisor: Torbj\"orn Sj\"ostrand}\\ [2mm]
{\it Department of Theoretical Physics,}\\[1mm]
{\it Lund University, Lund, Sweden}
\end{center}
 
\vspace{\fill}
\begin{center}
{\bf Abstract}\\[2ex]
\begin{minipage}{\abstwidth}
The transverse momentum of a colour-singlet massive particle in a hadronic collision is built up by two components, the perturbative effect of parton branchings and the nonperturbative effect of primordial $\kt$. In previous studies of transverse momentum spectra for $\mrm{Z}^0$ production at the Tevatron, the best fit to the experimental data are given when the primordial $\kt$ is set to a much higher value than what is expected considering the confinement of partons in the proton. We here investigate the possibility that the reason for this is that too few branchings are generated in showers, compared to the evolution used in the tuning of parton densities. This could then be compensated by increasing the value of $\lqcd$. The study is done using the regular $\py$ showering routines and a new algorithm where the branchings are ordered in transverse momentum $\pt^2$ instead of virtuality~$\q$.
\end{minipage}
\end{center}

\vspace{\fill}
 
\clearpage
\pagestyle{plain}
\setcounter{page}{1}
%

\section{Introduction}
\label{sec-intro}
\begin{quotation} 
\textit{Again, of bodies some are composite, others the elements of which these composite bodies are made up. These elements are indivisible and unchangeable, and necessarily so, if things are not all to be destroyed and pass into non-existence, but are to be strong enough to endure when the composite bodies are broken up, because they possess a solid nature and are incapable of being anywhere or anyhow dissolved. It follows that the first beginnings must be indivisible, corporeal entities.}
\begin{flushright}
Epicurus ``Letter to Herodotus''\\
approximately 300 B.C.
\end{flushright}
\end{quotation}

What are the building blocks of the universe? That is a very profound question, pondered upon by philosophers through the milleniums. For a long time the theories were purely metaphysical, with no possibility whatsoever to be tested experimentally. Thanks to major breakthroughs in theoretical and experimental physics during the twentieth century we now have the Standard Model, describing those elementary particles and their interactions. This model has been extremely successful in predicting the outcome of experiments, performed at increasingly large accelerators.

The elementary particles are divided into quarks, leptons and gauge bosons. Also the corresponding antiparticles exist. The gauge bosons mediate the four fundamental forces of nature: electromagnetism, the weak force, the strong force and gravity. The interesting particles for this thesis are the quarks and the gluons, the gauge bosons mediating the strong force. These are denoted q and g, with $\qbar$ denoting an antiquark. The theory describing the interactions of quarks and gluons is called quantum chromo dynamics, abbreviated QCD.

Free quarks have never been observed. The theoretical explanation for this has to do with the behaviour of the strong force. Here, it is useful to make a comparison with electromagnetism. Instead of electromagnetic charge, the strong force acts on particles with colour charge. This is, of course, not real colour, but is named this way since the behaviour of the colour charges resembles the behaviour of real colours in the way that mixing all the colour charges makes a colour neutral object. Of the elementary particles it is the quarks and gluons that carry colour charge. The crucial point is that the force between two colour charged particles does not decrease with the distance separating them, as is the case for electromagnetism. This means that it takes an infinite amount of energy to separate a q$\qbar$ pair, if they were to be connected to each other by a colour field. When the distance between the quarks gets sufficiently large, the potential energy becomes large enough to create a new q$\qbar$ pair. The new q$\qbar$ pair screens the colour field between the original charges. But now there are two colour connected q$\qbar$ pairs. So when the quark and the antiquark of a colour connected q$\qbar$ pair are moved away from each other, new q$\qbar$ pairs are created, but free quarks will not be obtained.

Another important difference from electromagnetism is that the particle mediating the strong force, the gluon, is not a colour neutral object. This means that a gluon can emit a gluon. So there are three different QCD branchings of particles: q~$\rightarrow$~qg and g~$\rightarrow$~q$\qbar$, which have their corresponding branchings in electromagnetism (q~$\rightarrow$~q$\gamma$ and $\gamma$~$\rightarrow$~q$\qbar$ respectively), and g~$\rightarrow$~gg, which has no similar branching in electromagnetism.

Quarks can carry the colour charges red, green or blue, the antiquarks the complementary anticolours and gluons one colour and one anticolour. As noted above only colour neutral objects can appear as free particles. Colour neutral particles out of quarks can be obtained in two different ways, either q$\qbar$ (colour-anticolour) or qqq (the mixture of red, green and blue makes a colour neutral object). The former particles are called mesons and the latter baryons (of course also antibaryons $\qbar\,\qbar\,\qbar$ exist), of which the proton and the neutron are the most well-known. The common name for all particles consisting of quarks is hadrons.

There are three generations of quarks $\left(\begin{tabular}{l} u \\ d \end{tabular} \right)$, $\left(\begin{tabular}{l} c \\ s \end{tabular} \right)$ and $\left(\begin{tabular}{l} t \\ b \end{tabular} \right)$. The u, d and s quarks are fairly light and will be assumed to be massless in this thesis, the c and b quarks are considerably heavier and the t quark is very heavy and too shortlived to be bound in any hadron.

To test the Standard Model experimentally, particles are accelerated to high energies and then collided. In the collisions the energy can be used to create new particles. In this way many particles predicted by the Standard Model have been found. Much of the effort in particle physics today is aimed at finding the Higgs particle, the last particle predicted by the Standard Model that has not yet been found. This search has been performed at the Tevatron, an accelerator at Fermilab outside of Chicago, and will be continued at LHC (Large Hadron Collider), an accelerator under construction at CERN in Geneva.

At the Tevatron protons and antiprotons are collided and at LHC protons are collided with protons. The energies are high enough to resolve the quarks and the gluons in the proton; these are called partons with a common name. The process of interest is often a hard collision between partons, which is not easy to extract from the experimental data or calculate theoretically. Firstly, in the final state, the partons created at the high energy scale of the hard collision branch into showers of partons. These parton showers evolve to a lower energy scale, where the partons are bound into hadrons. It is these hadrons that show up in the detector. The hadronization of partons can not be described by perturbative methods. Secondly, in the initial state, when the accelerated particles are protons, there is a problem of knowing the inner structure of the proton at different energy scales. High energy scales correspond to short distance interactions, as hard collisions of partons. Since the process of interest is a hard collision of partons, the distribution of partons in the proton at this energy scale must be known. The partons of the hard collision can be generated by letting the partons of the proton at a lower energy scale develop initial-state parton showers up to the scale of the hard collision. These initial-state parton showers will be the subject of the study in this thesis. 

At the scale of particle physics our universe is described by quantum mechanics, which means that it is not possible to predict the outcome of a single event. Only the probabilities of what is going to happen can be predicted. It is therefore necessary to average over a large number of events in a meaningful way. Perturbative methods can be applied to describe events down to a low energy scale, under which nonperturbative methods have to be used. In an event hundreds of particles can be produced. This means that even if the calculations could be done using perturbative methods, the complexity would exclude many approaches. One approach that does work is doing Monte Carlo simulations of a large number of events. Both initial-state and final-state parton showers are generated. This way spectra of different variables, that can be compared to experiment, are obtained. Event generators are used to interpret the outcome of experiments and to predict what to look for in the future. One of the most successful event generators, that will be used in this thesis, is called $\py$ \cite{Sjostrand:2000wi, Sjostrand:2003wg}.

One interesting variable to simulate in event generators is transverse momenta $\pt$, the momentum that particles have transverse to the accelerated beam. In the branchings the partons undergo, they get recoil transverse to the beam axis. This means that the more branchings, the more $\pt$ of the partons. However there is also an opposite effect, the accumulated $\pt$ of a parton is split between the two new partons of a branching. Still, the branchings affect $\pt$, so to look at $\pt$ is a good way to find out something about the branchings of an event. $\pt$ can be seen as a measure of how violent a process is. 

One process where it is possible to make a comparison between theory, event generators and experimental data in a rather clean way is q$\qbar$~$\rightarrow$~$\mrm{Z}^0$. When the $\mrm{Z}^0$ decays to an $\eplus\eminus$ pair, its $\pt$ is reasonably easy to measure. The $\pt$ of the $\mrm{Z}^0$ boson is the sum of the $\pt$ of the partons in the hard collision, where it was created. The $\pt$ of these partons is built up by two components; a perturbative component from the parton branchings and a non-perturbative component called primordial $\kt$ to be described as follows. The partons in the proton have an uncertainty in momenta that, according to Heisenberg's uncertainty relation, should be inversely proportional to the radius of the proton. This gives rise to a transverse momentum, that is called primordial $\kt$. For high $\pt$ values of the $\mrm{Z}^0$ boson the perturbative component from the parton branchings dominates, while the low $\pt$ part of the $\pt$-spectrum is sensitive to primordial $\kt$.

The $\pt$ of the $\mrm{Z}^0$ and $\mrm{W}^\pm$ bosons in the process q$\qbar$~$\rightarrow$~$\mrm{Z}^0$/$\mrm{W}^\pm$ have been measured at the Tevatron \cite{Abe:1991rk, Abbott:1998jy, Affolder:1999jh} and there is a problem with the $\py$ prediction of the $\pt$-spectrum for small $\pt$ as follows. To get the best fit to experimental data for small $\pt$ the value of the primordial $\kt$ has to be set considerably higher than expected, considering the radius of the proton. 

This thesis investigates one possible cause to this problem. The study will be made for the case of $\mrm{Z}^0$ production. In $\py$, some perturbative branchings, that normally are assumed to contribute, are not simulated. This means that the $\pt$-spectrum of $\mrm{Z}^0$ boson will be shifted to lower values of $\pt$. To compensate for this the other component building up the $\pt$ of the $\mrm{Z}^0$ boson, the primordial $\kt$, has to be increased. 

If the cause of the problem is that $\py$ generates too few branchings, there might be a way of solving it. In the expression for the running strong coupling constant $\as$($\q$) there is a parameter $\lqcd$ that has to be measured experimentally. An increase of $\lqcd$ increases the rate of branchings. At first sight this does not seem to provide any solution to the problem, since the same $\lqcd$ used to determine the parton densities, which are used in $\py$, must be used in the $\py$ parton showers. However, the parton densities are leading log. In $\py$ there are some corrections to pure leading log DGLAP evolution, introduced to give better agreement with next to leading log results. It is these corrections that decrease the rate of branchings, so to compensate for this $\lqcd$ can be increased. What is studied in this thesis is how much $\lqcd$ should be increased to compensate for the corrections and to what extent this solves the problem with primordial $\kt$. 

In Section~\ref{sec-psh} the concepts of parton showers and the DGLAP equations are introduced and a more specific description of the way of doing initial-state parton showers in $\py$ is given. Section~\ref{sec-prim} is a description of the problem with the too high value of primordial $\kt$, when comparing to experiment. The corrections to DGLAP evolution in $\py$ are described in Section~\ref{sec-corr}. The study of the generation of $\pt$-spectra is presented in Section~\ref{sec-study}. In this section also a new algorithm for doing initial-state showers is introduced. This new algorithm does not include heavy quarks. Section~\ref{sec-hq} is a attempt to find a way of introducing heavy quarks in the new algorithm. And finally Section~\ref{sec-sum} is a summary of the results of this thesis.

\section{Parton showers}
\label{sec-psh}
To simulate events, that can be compared with the data from accelerators, either matrix element calculations or parton showers can be used. In the matrix element approach amplitudes and phases are used to do a fully quantum mechanical calculation. Calculating matrix elements order by order in the strong coupling constant $\as$ is a good, but complicated, method to get cross sections for processes. As each branching contributes with one order in $\as$, the matrix element calculations get too complicated for processes involving more than a few branchings. An approximate approach is needed. In such an approach probability distributions, instead of amplitudes and phases, can be used. This is an ideal method for doing computer simulations. Using the probabilities for the different branchings, given by the DGLAP equations \cite{DGLAP}, one parton branching into a shower of partons can be simulated \cite{Webber:mc}. This method can be applied to events with arbitrary many branchings, but is only accurate to what is called leading log, where to each order in $\as$, only the most divergent contribution is described correctly. In this parton shower approach the histories of the partons are given, so they can be traced back through the branchings. 

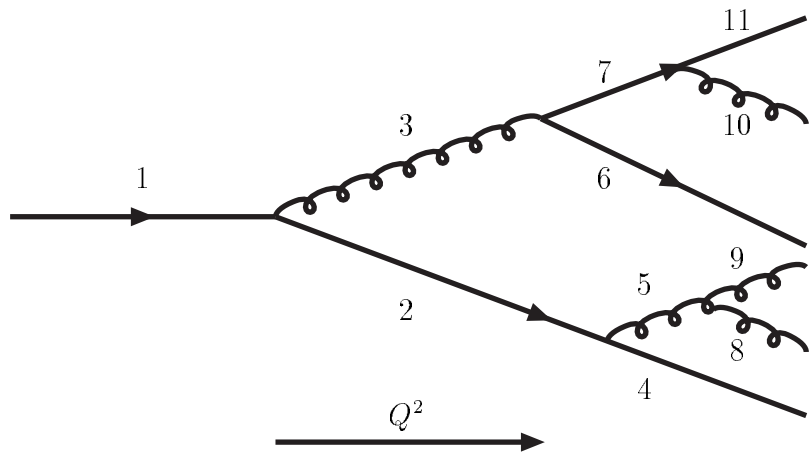
\begin{figure}[t]
\begin{center}
\begin{picture}(300,170)(0,0)
\SetWidth{2}
\ArrowLine(0,95)(100,95)
\ArrowLine(100,95)(300,20)
\ArrowLine(200,132)(300,170)
\ArrowLine(200,132)(300,84)
\Gluon(100,95)(200,132){3}{7}
\Gluon(225,48)(300,76){3}{5}
\Gluon(265,60)(300,44){3}{2}
\Gluon(250,150)(300,130){3}{3}
\LongArrow(100,10)(200,10)
\Text(150,20)[]{$\q$}
\Text(50,110)[]{1}
\Text(150,60)[]{2}
\Text(150,130)[]{3}
\Text(240,30)[]{4}
\Text(240,70)[]{5}
\Text(225,110)[]{6}
\Text(225,150)[]{7}
\Text(275,45)[]{8}
\Text(275,80)[]{9}
\Text(275,130)[]{10}
\Text(275,170)[]{11}
\end{picture}
\end{center}
\caption{From the three processes q~$\rightarrow$~qg, g~$\rightarrow$~gg and g~$\rightarrow$~q$\qbar$ it is possible to build up complicated showers. The emissions are ordered in $\q$.}
\label{fig:part}
\end{figure}    
%

\subsection{General concepts of parton showers}
\label{subsec-gen}
In the introduction the proton is said to consist of three quarks; this is a simplification. In the proton different processes is constantly occuring: q~$\rightarrow$~qg, g~$\rightarrow$~gg, g~$\rightarrow$~q$\qbar$ and their inverses. The quantum numbers of a quark originating from a gluon splitting are always compensated by the quantum numbers of the antiquark originating from the same splitting. These quarks are called sea quarks, while the three quarks that determine the quantum numbers of the proton are called valence quarks in analogy with atomic physics.  

As a consequence of these constantly occuring processes, the structure of the proton depends on the momentum scale $\q$ it is probed by. A higher $\q$ can resolve more gluons and sea quarks. To describe the structure of the proton, parton density functions $f_a(x,\q)$ are used. They express the probability of finding a parton of type $a$ with a fraction $x$ of the total proton momentum in a proton probed with $\q$. To calculate parton density functions nonperturbative input is needed. This nonperturbative input is equivalent to having the $f_a(x,\q)$ defined at one $\q = \q_0$. Given such initial conditions, the parton density function at another $\q$ can be calculated, see Subsection~\ref{subsec-DGLAP}. The $\q$ dependance of the parton density functions is often called scaling violations, since the $f_a$ are no longer functions of $x$ alone.

Another way of describing this gradual resolution of the proton is to start with one initial parton and do an evolution in $\q$. As $\q$ increases the parton branches into more and more partons and a shower of partons arises. Such a parton shower is built up by the three processes (if processes involving photons are neglected) q~$\rightarrow$~qg, g~$\rightarrow$~gg and g~$\rightarrow$~q$\qbar$, see Fig.~\ref{fig:part}. To do a simulation of parton showers, the probability for these processes as a function of $\q$ is needed. In Subsection~\ref{subsec-DGLAP} the DGLAP equations, giving these probabilities, are introduced.

This parton shower approach can be used to generate events at particle colliders. An event is separated into one hard process, with the highest $\q$, and many softer interactions both before and after the hard process. To describe these softer interactions, the parton shower approach is useful. The parton showers of an event can be separated into final-state showers and initial-state showers. The final-state showers evolve if the final particles of the collision are quarks or gluons. This can be simulated by an evolution in $\q$, starting with the quarks at the scale of the hard collision and going down to the scale of hadronization. If the colliding particles are protons there is also initial-state parton showers, starting with the partons at some low scale $Q_0^2$ going up to the scale of the hard collision. In Fig.~\ref{fig:part} the parton shower starts with a quark 1 at $Q_0^2$, which branches into a quark 2 and a gluon 3 at a higher $\q$. At an even higher $\q$ quark 2 branches into a quark 4 and gluon 5 and at another $\q$ gluon 3 splits into a quark 6 and an antiquark 7. The branchings continue up to the scale of the hard collision. Only one of the partons in the cascade takes part in the hard collision. When the partons that participates in the hard collisions are given, they can be traced back to the low scale by backwards evolution, to be described below. The backwards evolution starts with the parton of the hard collision, for instance gluon 8, tracing it to the branching of gluon 5 at a lower $\q$. Gluon 5 is emitted from quark 2 at an even lower $\q$. Quark 2 originates from the branching of quark 1, which is the initial parton of the shower at scale $Q_0^2$. In this way the $\pt$ of the colliding particles, and consequently the $\pt$ of the produced particle, can be calculated. 

\subsection{DGLAP equations}
\label{subsec-DGLAP}

In DGLAP evolution the emissions are ordered in spacelike virtualities $\q$ and the probability of the process $a \rightarrow bc$ is given by

\begin{equation}
\dint \mathcal{P}_{a \rightarrow bc} = \frac{\dint \q}{\q} \frac{\as(\q)}{2 \pi} P_{a \rightarrow bc}(z)
\label{eq:DGLAP}
\end{equation}  
where $P_{a \rightarrow bc}(z)$ is the splitting function depending on $z$, the fraction of the momentum of particle $a$ taken by particle $b$. For the processes that are used to build up a QCD shower, the splitting functions are given by

\begin{eqnarray}
P_{\mrm{q} \rightarrow \mrm{qg}}(z) &=& \frac{4}{3} \frac{z^2 + 1}{1 - z} \label{eq:spl}\\
P_{\mrm{g} \rightarrow \mrm{gg}}(z) &=& 3 \frac{(1-z(1-z))^2}{z(1 - z)} \label{eq:spli}\\
P_{\mrm{g} \rightarrow \mrm{q\qbar}}(z) &=& \frac{1}{2} (z^2 + (1 - z)^2)
\end{eqnarray}

In eq.~(\ref{eq:DGLAP}) the running strong coupling constant is used
 
\begin{equation}
\as(\q) = \frac{12 \pi}{(33 - 2 n_{\mrm{f}}) \ln \left(\frac{\q}{\lqcd}\right)}
\label{eq:as}
\end{equation} 
where $n_{\mrm{f}}$ is the number of flavours and $\lqcd$ is a parameter of the Standard Model that has to be measured experimentally. $\lqcd$ will be of great importance for this thesis. 

The parton shower approach is an exclusive approach where one parton, that branches into a shower as $\q$ increases, is studied. To include all the partons of the proton, one can study the behaviour of parton densities as $\q$ increases. Beginning with initial conditions for the parton densities at one $\q_0$, considering the evolution of all the partons, the parton densities at another $\q$ can be calculated. The dependence on $\q$ of the parton density for a particle of type $b$ at the value $x$ is given by  

\begin{equation}
\frac{\dint \, f_b(x,\q)}{\dint \, (\ln \q)} = \sum_a \int_x^1 \frac{\dint y}{y} f_a(y,\q) \, \frac{\as}{2 \pi} P_{a \rightarrow bc}\left(\frac{x}{y}\right)
\label{eq:AP}
\end{equation} 

\subsection{Initial-state showers in $\py$}
\label{subsec-inp}
In this thesis initial-state showers in $\py$ \cite{Sjostrand:1985xi, Bengtsson:1986gz, Miu:1998ju} are studied. $\py$ uses the parton shower approach to do Monte Carlo simulations of events. The DGLAP equations are used to simulate at which $\q$, and with which $z$, the branchings occur.

In the evolution a branching can only occur if the parton has not already branched. This introduces a Sudakov form factor to the expression for the probability

\begin{equation}
\dint \mathcal{P}_{a \rightarrow bc} = \frac{\dint \q}{\q} \frac{\as(\q)}{2 \pi} P_{a \rightarrow bc}(z) \exp \left(-\int_{\q_0}^{\q} \frac{\dint Q^{\,\prime 2}}{Q^{\,\prime 2}} \frac{\as(Q^{\,\prime 2})}{2 \pi} P_{a \rightarrow bc}(z) \right) 
\label{eq:sud}
\end{equation}

To do a Monte Carlo simulation, a function $N(\q)$ for the number of partons that has not branched is needed 

\begin{equation}
N(\q) = N(\q_0)(1 - \int_{\q_0}^{\q} \dint \mathcal{P})
\end{equation}
This number is then set to a random number $R$ times the number of partons at $\q_0$, $N(\q_0)$. So the $\q$ at which the branching occurs is given by

\begin{eqnarray}
\int_{\q_0}^{\q} \dint \mathcal{P} &=& 1 - R = R \nonumber \\
&\Rightarrow& \exp \left(-\int_{\q}^{\q_{\mrm{max}}} \frac{\dint Q^{\,\prime 2}}{Q^{\,\prime 2}} \frac{\as(Q^{\,\prime 2})}{2 \pi} \int \dint z P_{a \rightarrow bc}(z) \right) = R 
\end{eqnarray}
using the fact that $1 - R$ is a random number between 0 and 1 if $R$ is it. The $z$ of the branching can then be generated by setting 

\begin{equation}
\int_0^z \dint z^{\prime} P_{a \rightarrow bc}(z^{\prime}) = R \int_0^1 \dint z^{\prime} P_{a \rightarrow bc}(z^{\prime})
\label{eq:z}
\end{equation}
As seen in eq.~(\ref{eq:spl}) and eq.~(\ref{eq:spli}) the probability of emissions of very soft gluons, where the gluon takes a very small part of the momentum, goes to infinity. It is therefore necessary to introduce a cut-off $\epsilon$ in $z$. In the case of $\mrm{q} \rightarrow \mrm{qg}$ only an upper cut-off is needed, while in the case of $\mrm{g} \rightarrow \mrm{gg}$ both an upper and a lower cut-off is needed. Instead of integrating from 0 on the left side and up to 1 on the right side in eq.~(\ref{eq:z}), the integration is from $\epsilon$ and up to 1 - $\epsilon$ respectively. 

In the forward evolution $\q$ is generated this way for the different possible branchings (q~$\rightarrow$~qg, g~$\rightarrow$~gg and g~$\rightarrow$~q$\qbar$) and the one with the lowest $\q$ is chosen. Forward evolution can be very inefficient when a specific kind of hard process is studied, since the final-state partons are not known beforehand. If no final-state partons with the right properties to participate in the hard process are generated in the evolution, it can not be used. A more efficient method, used in $\py$, is to fix the hard scattering using evolved parton densities and thereafter use backwards evolution to form this into exclusive events. Such an approach can again be formulated in Monte Carlo terms, but now starting from eq.~(\ref{eq:AP}) instead of eq.~(\ref{eq:DGLAP}). This gives

\begin{equation}
\exp \left(-\int_{\q}^{\q_{\mrm{max}}} \frac{\dint Q^{\,\prime 2}}{Q^{\,\prime 2}} \frac{\as(Q^{\,\prime 2})}{2 \pi} \int_x^1 \frac{\dint y}{y} \frac{f_a(y,\q)}{f_b(x,\q)} P_{a \rightarrow bc} \left( \frac{x}{y} \right) \right) = R 
\end{equation}
where $z$ has been rewritten as $x$, the momentum fraction of the parton before (in the ordering of the backwards evolution) the branching, divided by $y$, the momentum fraction of the parton after the branching. In the exponential the probability is integrated from $\q$ to $\q_{\mrm{max}}$, the virtuality of the preceding step, instead of from $\q_0$ to $\q$, as is the case for forward evolution.

The same procedure can then be repeated starting at the $\q$ of the branching. In this way the partons of the hard interaction are evolved back to the initial partons at the lower scale $\q_0$.

\section{Primordial $\kt$}
\label{sec-prim}
The partons are confined in the proton of radius $r_{\mrm{p}} \approx \mrm{0.7\,fm}$. According to Heisenberg's uncertainty relation this means that they have a momentum inside the proton. The part of this momentum that is in the beam direction gives the fraction $x$ of the total proton momentum carried by the parton, described by parton density functions. But there is also a part of this momentum that is transverse to the beam direction. This transverse momentum of the partons, due to the confinement in the proton, is called primordial $\kt$ and its mean value can be estimated by

\begin{equation}
\langle \kt \rangle\approx\frac{\hbar}{r_{\mrm{p}}}\approx\frac{\mrm{0.2\,GeV \cdot fm}}{\mrm{0.7\,fm}}\approx\,\mrm{0.3\,GeV}
\label{eq:heis}
\end{equation}

\begin{figure}[t]
\begin{center}
\begin{picture}(400,150)(0,0)
\SetWidth{2}
\LongArrow(50,5)(50,25)
\LongArrow(50,145)(50,125)
\LongArrow(180,146)(180,160)
\LongArrow(180,144)(180,130)
\LongArrow(230,21)(230,35)
\LongArrow(230,19)(230,5)
\LongArrow(240,121)(240,135)
\LongArrow(240,119)(240,105)
\LongArrow(290,46)(290,65)
\LongArrow(290,44)(290,25)
\ArrowLine(50,150)(100,145)
\ArrowLine(50,0)(150,20)
\ArrowLine(100,145)(200,120)
\ArrowLine(150,20)(250,45)
\ArrowLine(200,120)(300,75)
\ArrowLine(250,45)(300,75)
\Gluon(100,145)(200,165){3}{8}
\Gluon(150,20)(230,0){3}{5}
\Gluon(200,120)(300,165){3}{8}
\Gluon(250,45)(325,0){3}{7}
\Photon(300,75)(400,75){3}{8}
\Text(40,140)[r]{Primordial $\kt$}
\Text(40,10)[r]{Primordial $\kt$}
\Text(350,85)[]{$\mrm{Z}^0$}
\end{picture}
\end{center}
\caption{Production of a $\mrm{Z}^0$-boson in a parton shower model. At each branching the accumulated $\pt$ is split between the particles in $z$ vs. $1 - z$ fractions. New $\pt$ values for the partons after the branching are generated, and the shower continues. This means that showers with many branchings are not very sensitive to the value of primordial $\kt$.}
\label{fig:Z-prod}
\end{figure}
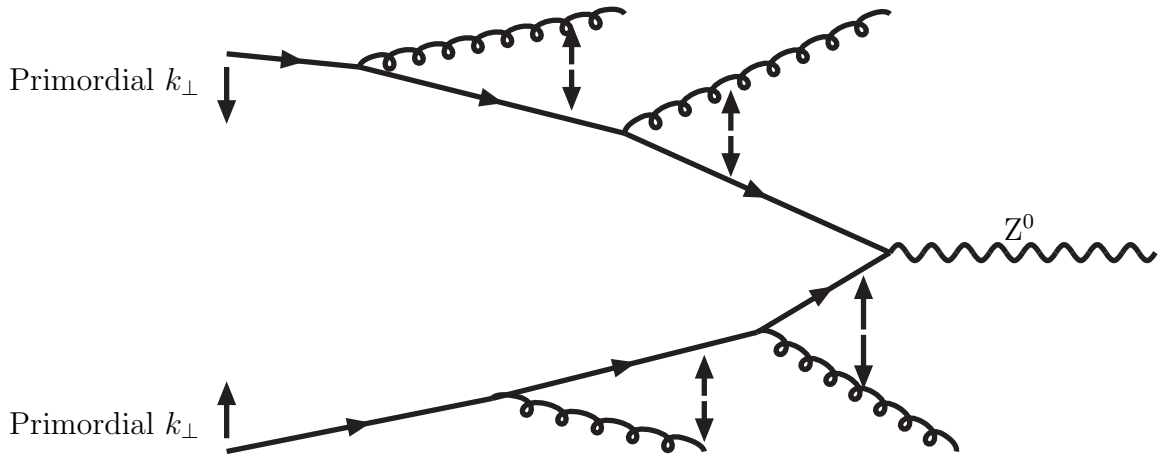

To compare this value to experiment, we can look at the $\pt$-spectrum of the produced $\mrm{Z}^0$ boson in the process q$\qbar$~$\rightarrow$~$\mrm{Z}^0$. The reason for looking at this particular process is that when the $\mrm{Z}^0$ boson decays to a $\eplus\eminus$ pair the $\pt$ of the $\mrm{Z}^0$ boson is reasonably easy to measure. The $\pt$-spectrum of the $\mrm{Z}^0$ boson depends on the value of the primordial $\kt$ and on the $\pt$-kicks the quarks get when they emit gluons. For high $\pt$ values of the $\mrm{Z}^0$ boson the effect of the $\pt$-kicks in the shower dominates, while the low $\pt$ part of the $\pt$-spectrum is sensitive to primordial $\kt$. Fig.~\ref{fig:Z-prod} gives a schematical picture of this. In one set of data from the Tevatron the $\py$ prediction of the $\pt$-spectrum fits the experimental data best for a primordial $\kt$ value of 2.15~Gev \cite{Balazs:2000sz}. This value is considerably higher than the value estimated by Heisenberg's uncertainty relation. One reason for this could be that $\py$ has too low a rate of emissions. Then there would be too few $\pt$-kicks and the value of primordial $\kt$ would have to be raised to increase the $\pt$ values of the $\mrm{Z}^0$-boson to the level of the experimental data. Another reason could be that soft interactions that fall below the cut-off in $\pt$ can build up a $\pt$ value of the partons before the parton shower starts.  

One specific possibility for too low a rate of emissions in $\py$, explored in this thesis, is the following. $\py$ version 6.220 uses CTEQ5L parton densities \cite{CTEQ5L}. These are leading log, determined with a $\lqcd$ value of 0.192~Gev. However, the parton showers in $\py$ are not pure leading log. They include corrections to simple DGLAP evolution in $\q$, that decrease the rate of the $x$ evolution, so that the $x$ evolution is slower than in the measurement of the parton densities. This could be compensated by raising the value of $\lqcd$. The higher $\lqcd$ value gives more emissions, which means that a lower value of primordial $\kt$ is needed for $\py$ to give a $\pt$-spectrum similar to the experimental one. This could solve the problem with the disagreement between the primordial $\kt$ value estimated by Heisenberg's uncertainty relation and the experimental value. Which corrections $\py$ includes is described in Section~\ref{sec-corr}.

It should also be noted that the primordial $\kt$ of the initial partons is split between the partons in each branching of the parton shower. This means that the partons of the hard collision only carry a fraction of the primordial $\kt$ of the initial partons

\begin{equation}
(\kt)_{\mrm{at \, hard \, int}} \approx \frac{\langle x_{\mrm{hard}} \rangle}{\langle x_{\mrm{initial \, parton}} \rangle} (\kt)_{\mrm{initial \, parton}}
\label{eq:kt}
\end{equation}

%

\section{Corrections to pure DGLAP evolution in $\py$}
\label{sec-corr}
There are a number of corrections to pure leading log DGLAP backwards evolution in $\py$. These, and their effect on the rate of emissions, will be described in this section. 

\subsection{Angular ordering}
\label{subsec-ang}
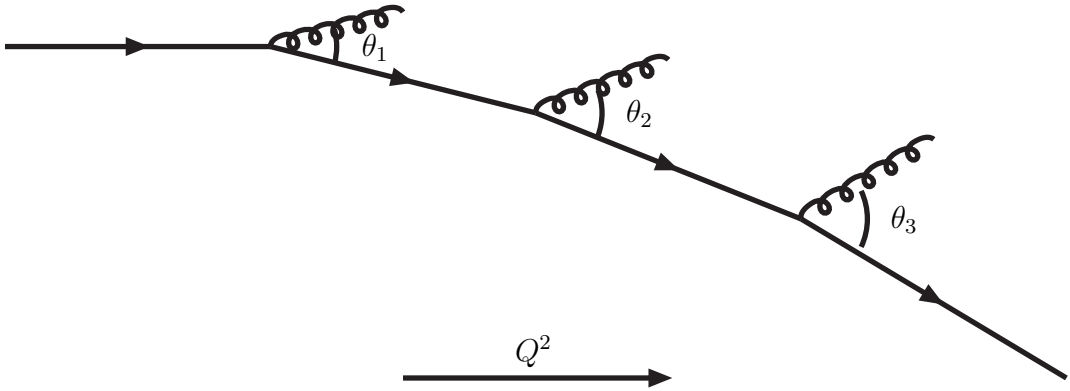
\begin{figure}[t]
\begin{center}
\begin{picture}(400,138)(0,0)
\SetWidth{2}
\LongArrow(150,0)(250,0)
\ArrowLine(0,125)(100,125)
\ArrowLine(100,125)(200,100)
\ArrowLine(200,100)(300,60)
\ArrowLine(300,60)(400,0)
\Gluon(100,125)(150,138){3}{5}
\Gluon(200,100)(250,120){3}{5}
\Gluon(300,60)(350,90){3}{5}
\CArc(100,125)(25,-15,15)
\CArc(200,100)(25,-20,20)
\CArc(300,60)(25,-25,25)
\Text(200,10)[]{$Q^\mrm{2}$}
\Text(135,125)[l]{$\theta_\mrm{1}$}
\Text(235,100)[l]{$\theta_\mrm{2}$}
\Text(335,60)[l]{$\theta_\mrm{3}$}
\end{picture}
\end{center}
\caption{Emissions in $\py$ are angular ordered: $\theta_\mrm{1}\,<\,\theta_\mrm{2}\,<\,\theta_\mrm{3}$.}
\label{fig:ang}
\end{figure}

Due to coherence effects the opening angles of emitted partons increase with increasing virtuality $\q$, see Fig.~\ref{fig:ang}. The approximate opening angles can be calculated using $x$, $z$, $\q$ and $s$

\begin{equation}
\theta^2 \approx \frac{4 z^2 \q}{4 z^2 \q +(1 - z) x^2 s}
\label{eq:ang}
\end{equation}
If an emission has a larger opening angle than the previous one in the $\py$ backwards evolution, the emission is rejected. This reduces the rate of emissions.
%
\subsection{The condition that $\uhat\,<\,0$}
\label{subsec-uhat}
\begin{figure}[t]
\begin{center}
\begin{picture}(200,170)(0,0)
\SetWidth{2}
\ArrowLine(0,150)(100,90)
\ArrowLine(100,90)(150,50)
\ArrowLine(0,0)(150,50)
\Gluon(100,90)(150,130){3}{5}
\Photon(150,50)(200,50){2}{5}
\Text(175,60)[]{$\mrm{Z}^0$}
\Text(140,75)[]{$\q$}
\Text(90,85)[]{$z$}
\Text(175,40)[]{$\mz$}
\Text(0,160)[l]{a)}
\end{picture}
\begin{picture}(200,170)(0,0)
\SetWidth{2}
\ArrowLine(0,150)(100,100)
\ArrowLine(100,100)(100,50)
\ArrowLine(0,0)(100,50)
\Gluon(100,100)(200,150){3}{8}
\Photon(100,50)(200,0){2}{8}
\Text(165,35)[]{$\mrm{Z}^0$}
\Text(110,75)[l]{$\that$}
\Text(50,75)[]{$\shat$}
\LongArrow(50,85)(50,120)
\LongArrow(50,65)(50,30)
\Text(150,10)[]{$\mz$}
\Text(0,160)[l]{b)}
\Text(20,150)[l]{1}
\Text(20,0)[l]{2}
\Text(175,0)[l]{4}
\Text(175,150)[l]{3}
\end{picture}
\caption{The emission of a gluon by one of the quarks at $\mrm{Z}^0$-production seen a) in a parton shower model and b) as a 2~$\rightarrow$~2 process.}
\label{fig:uhat}
\end{center}
\end{figure}
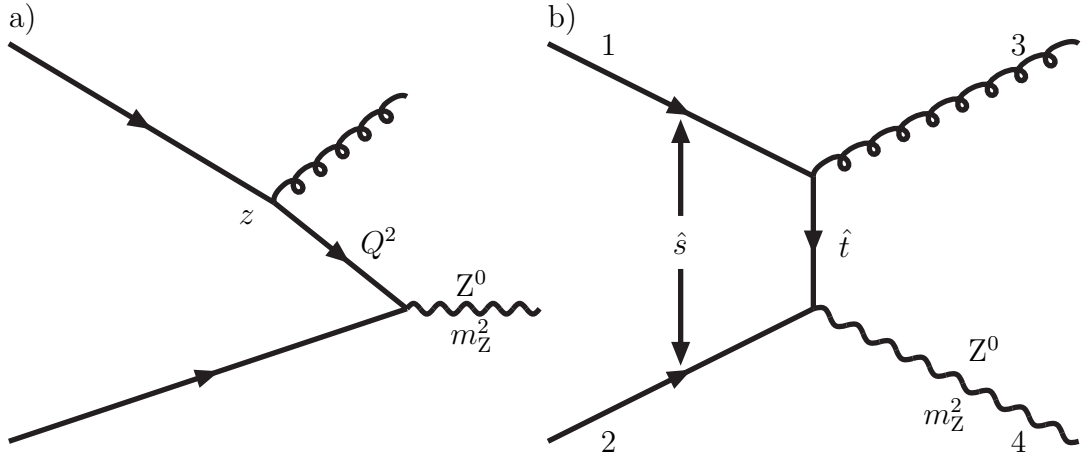

When one of the quarks in the process q$\qbar$~$\rightarrow$~$\mrm{Z}^0$ emits a gluon this can be seen both in a parton shower model, Fig.~\ref{fig:uhat}a, and as a 2~$\rightarrow$~2 process, Fig.~\ref{fig:uhat}b. The Mandelstam variables, Lorentz invariants of the kinematics, can be expressed in terms of the variables of the parton shower

\begin{eqnarray}
\shat &=& (P_1 + P_2)^2 = \frac{\mz}{z}\\
\that &=& (P_1 - P_3)^2 = - \q\\
\uhat &=& (P_1 - P_4)^2 =\sum \m - \shat - \that = \mz - \frac{\mz}{z} + \q = \q - \mz \frac{1-z}{z}
\label{eq:uhat}
\end{eqnarray}

In the picture of a 2~$\rightarrow$~2 process $\uhat\,<\,0$. For that reason in $\py$ there is a condition on the emissions that $\uhat\,<\,0$, using the final expression for $\uhat$ in eq.~(\ref{eq:uhat}). However, if the quark emitting the gluon already before th gluon emission has a negative $\m$ because of preceding shower evolution, the kinematical constraints on $\uhat$ are relaxed and $\uhat\,<\,0$ is just an assumption. No such kinematical constraints are present in the leading log DGLAP formalism. This means that the rate of emissions in $\py$ is lower than in pure DGLAP evolution.

%

\subsection{Evolution with $\as$($\pt^2$) instead of $\as$($\q$)} 
\label{subsec-apt}

The evolution in $\py$ is done with $\as$($\pt^2$) instead of $\as$($\q$). To examine the effect of this, an approximate expression for $\pt^2$ can be derived using the lightcone variables
\begin{eqnarray}
\pplus &=& E + \pz\\
\nonumber \\
\pminus &=& E - \pz
\end{eqnarray}
fulfilling
\begin{equation}
\pplus \cdot \pminus = E^\mrm{2} - \pz^\mrm{2} = \m + \pt^\mrm{2} 
\end{equation}
%
\begin{figure}[t]
\begin{center}
\begin{picture}(300,155)(0,0)
\SetWidth{2}
\ArrowLine(0,75)(150,75)
\ArrowLine(150,75)(300,0)
\Gluon(150,75)(300,150){3}{10}
\Text(130,40)[l]{$\m = - \q$}
\Text(130,20)[l]{$\pplus = z \cdot \pplus^\mrm{0}$}
\Text(130,0)[l]{$\pminus = \frac{- \q + \pt^\mrm{2}}{z \cdot \pplus^\mrm{0}}$}
\Text(25,130)[l]{$\m = 0$}
\Text(25,110)[l]{$\pplus = \pplus^\mrm{0}$}
\Text(25,90)[l]{$\pminus = 0$}
\Text(130,155)[l]{$\m = 0$}
\Text(130,135)[l]{$\pplus = (1 - z) \cdot \pplus^\mrm{0}$}
\Text(130,115)[l]{$\pminus = \frac{\pt^\mrm{2}}{(1 - z) \cdot \pplus^\mrm{0}}$}
\end{picture}
\end{center}
\caption{Emission of a gluon by a massless quark.}
\label{fig:lc}
\end{figure}    
%
The kinematics used in $\py$ is not this simple but the qualitative result can still be derived using lightcone variables. In Fig.~\ref{fig:lc} these variables are given for the case when a massless quark emits a gluon. Conservation of $\pminus$ gives
\begin{eqnarray}
0 &=& \frac{- \q + \pt^\mrm{2}}{z \pplus^\mrm{0}} + \frac{\pt^\mrm{2}}{(1-z) \pplus^\mrm{0}}  = \frac{- \q}{z \pplus^\mrm{0}} + \frac{\pt^\mrm{2}}{z (1-z) \pplus^\mrm{0}} \nonumber\\
\nonumber\\
&\Rightarrow& \pt^\mrm{2} = \q (1-z) 
\label{eq:pt}
\end{eqnarray}
So $\pt^\mrm{2} < \q$, which means $\as(\pt^2) > \as(\q)$ and thereby more emissions. This is not the whole picture, however. As seen in eq.~(\ref{eq:as}) $\as$($\q$) has a singularity when $\q = \lqcd$. To avoid this problem, in $\py$ there is a lower cut-off $Q_\mrm{0} = 1\, \mrm{GeV}$. The region below some cut-off of this order can not be described by perturbative methods. In $\py$ this region is instead accounted for by non-perturbative methods, such as primordial $\kt$. In the case of $\as$($\pt^2$) this lower cut-off must be applied to $\pt$ instead of $Q$, so that $\pt^2 > Q_\mrm{0}^\mrm{2}$. As seen in eq.~(\ref{eq:pt}) $\pt^2$ can be below the cut-off for high values of $\q$ if $z$ is sufficiently close to 1. This means that emissions with high values of $z$ are discarded. So the evolution with $\as$($\pt^2$), instead of $\as$($\q$), has two opposite effects on the rate of emissions. When studied the net effect is a decrease, see Subsection~\ref{subsubsec-pst}.

\subsection{Final-state showers of emitted particles with timelike virtuality} 
\label{subsec-fin}
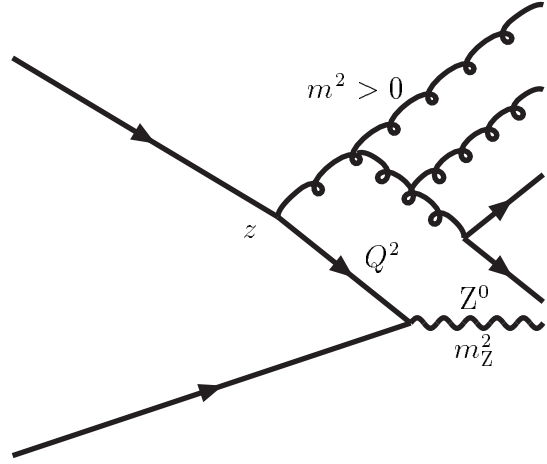
\begin{figure}
\begin{center}
\begin{picture}(200,170)(0,0)
\SetWidth{2}
\ArrowLine(0,150)(100,90)
\ArrowLine(100,90)(150,50)
\ArrowLine(0,0)(150,50)
\Gluon(100,90)(200,170){3}{6}
\Gluon(130,114)(170,82){3}{3}
\Gluon(150,98)(200,138){3}{4}
\Photon(150,50)(200,50){2}{5}
\ArrowLine(170,82)(200,58)
\ArrowLine(170,82)(200,106)
\Text(175,60)[]{$\mrm{Z}^0$}
\Text(140,75)[]{$\q$}
\Text(90,85)[]{$z$}
\Text(175,40)[]{$\mz$}
\Text(130,140)[]{$\m > 0$}
\end{picture}
\caption{An emitted gluon that develops a final-state shower corresponds to the gluon having $\m > 0$.}
\label{fig:time}
\end{center}
\end{figure}
Emitted particles that develop final-state showers of their own corresponds to particles with timelike virtuality ($\m > 0$ for gluons, $\m > \m_{\mrm{q}}$ for quarks), see Fig.~\ref{fig:time} . This does not affect the rate of emission but can affect the $\pt$-spectrum for the produced particle. To see this it is again convenient to use lightcone variables. If the gluon in Fig.~\ref{fig:lc} has a positive $\m$, eq.~(\ref{eq:pt}) becomes

\begin{eqnarray}
0 &=& \frac{- \q + \pt^\mrm{2}}{z \pplus^\mrm{0}} + \frac{\m + \pt^\mrm{2}}{(1-z) \pplus^\mrm{0}}  = \frac{- \q}{z \pplus^\mrm{0}} + \frac{\m}{(1-z) \pplus^\mrm{0}} + \frac{\pt^\mrm{2}}{z (1-z) \pplus^\mrm{0}} \nonumber\\
\nonumber\\
\nonumber \\
&\Rightarrow& \pt^\mrm{2} = \q (1-z) - z \, \m
\label{eq:ptm}
\end{eqnarray} 
The introduction of a positive $\m$ of the gluon clearly causes a decrease in $\pt$.

\subsection{Other corrections} 
\label{subsec-oth}
There are some other corrections that do not measurably affect either the rate of emissions or the $\pt$-spectrum for the produced particle very much. 

The probability of a quark emitting a gluon goes to infinity when the fraction of the momentum that the gluon gets in the emission, $1 - z$, goes to zero. To avoid this problem, in $\py$ there is an upper cut-off in $z$. In the case of a gluon emitting a gluon, the singularity occurs both at $z = 0$ and $z = 1$, and both a lower and an upper cut-off is needed. The very soft gluons that fall below this cut-off are resummed. 

Sometimes, e.g. when heavy quarks are produced, the reconstruction of the kinematics of a branching fails. The shower is then regenerated. This could affect the rate of emissions. When studied however, it does not seem to have any visible effect.

At large $\q$ the evolution is matched to first-order matrix elements. This is a separate issue from the parton density evolution and will not affect the discussion in this thesis.

Since quarks carry electric charge they can also emit photons. This is an electromagnetic interaction and thereby much less probable than the emission of a gluon. When studied, photon emissions do not seem to have any effect on the rate of the $x$ evolution either.  

\section{Study of $\lqcd$ values, $\pt$-spectra and primordial $\kt$}
\label{sec-study}
First the increase in $\lqcd$ needed to compensate for the corrections to pure DGLAP evolution will be quantified in this section. Then the effect of the increase in $\lqcd$ on $\pt$-spectra will be studied. The hope of this study was to be able to decrease the value of primordial $\kt$. This does not seem to be possible, so another approach is taken, the effect of final-state showers off emitted particles on $\pt$-spectra. Also a new shower algorithm is introduced. 

\subsection{Study of the increase in $\lqcd$ needed to compensate for the corrections to DGLAP} 
\label{subsec-lqcd}
\begin{figure}[t]
\begin{center}
\mbox{\epsfig{file=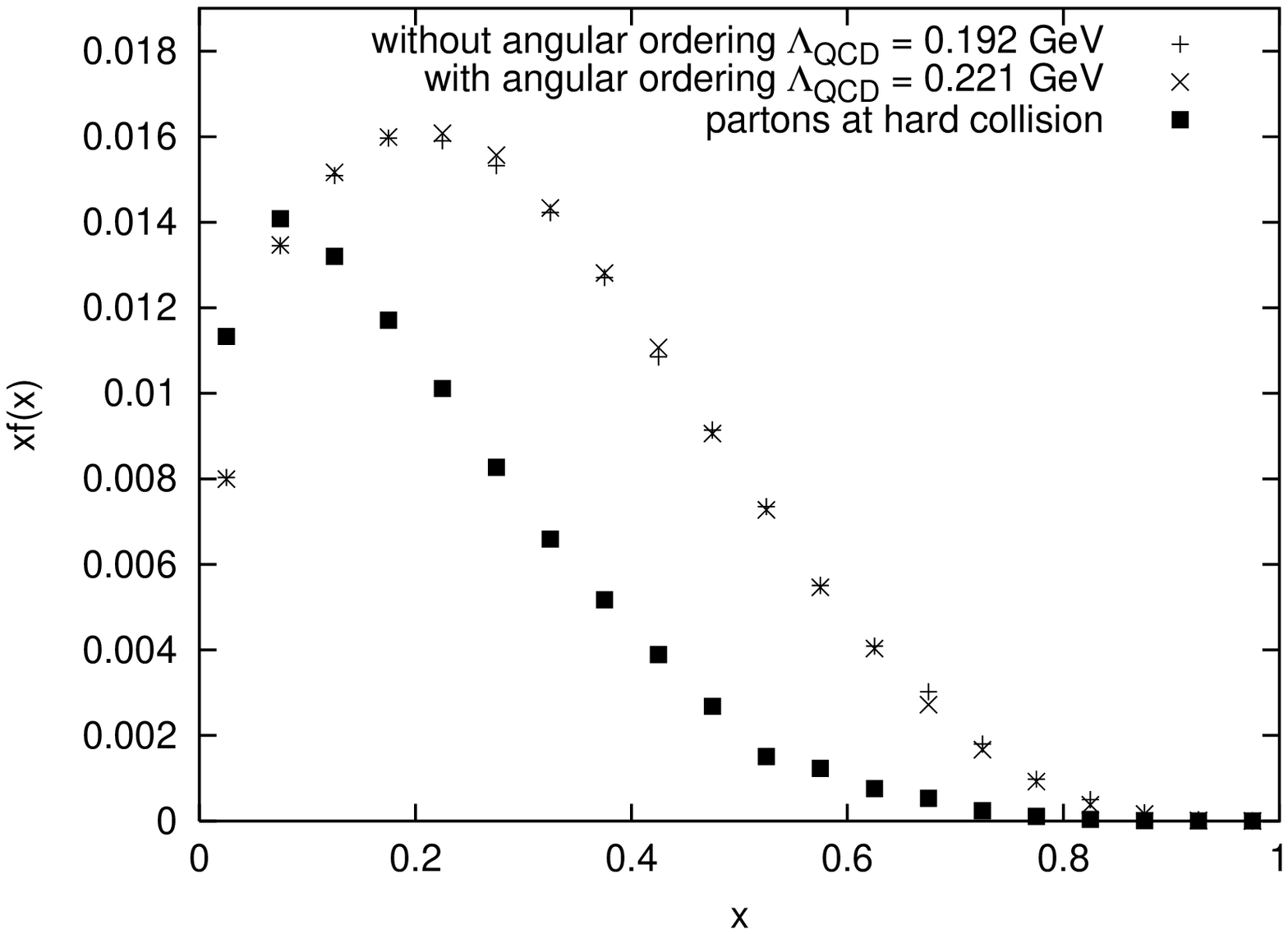,width=15cm}}
\caption{The two $x$-spectra generated by the toy model, with angular ordering using $\lqcd = 0.221 \,\mrm{GeV}$ and without angular ordering using $\lqcd = 0.192 \,\mrm{GeV}$ respectively, look close to identical. The $x$-spectra are generated by backwards evolution from the scale of the hard collision, $\q = \mz$, with a center of mass energy of 1800~Gev. For comparison, the $x$-spectrum for the partons at the hard collision is shown.}
\label{fig:comp}
\end{center}
\end{figure}

\begin{figure}[t]
\begin{center}
\mbox{\epsfig{file=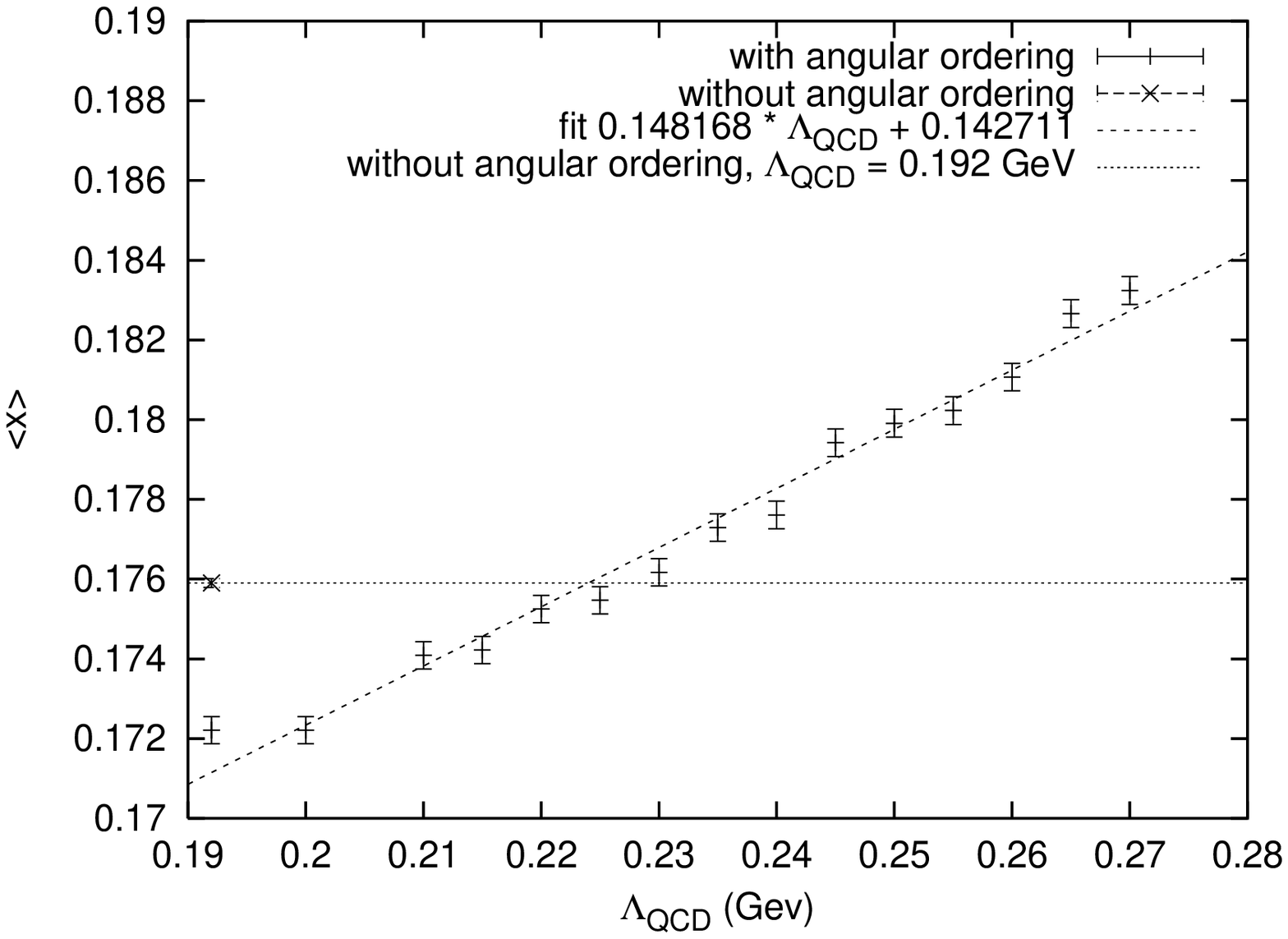,width=15cm}}
\caption{A way of quantifying the increase in $\lqcd$ needed to compensate for angular ordering for the process q$\qbar$~$\rightarrow$~$\mrm{Z}^0$ at the Tevatron. First the $x$-spectrum for the initial partons is generated without angular ordering using $\lqcd = 0.192 \,\mrm{GeV}$. Then $x$-spectra for the initial partons are generated with angular ordering using different values of $\lqcd$. At $\lqcd = 0.224 \,\mrm{GeV}$ the mean value of $x$ is the same as without angular ordering at $\lqcd = 0.192 \,\mrm{GeV}$.}
\label{fig:tevz}
\end{center}
\end{figure}

\subsubsection{Toy model studies} 
\label{subsubsec-toy}

Some of the main features can be studied in a toy model framework, using only the DGLAP equations to generate parton showers. In this framework a certain feature can be studied more directly, as opposed to the more complicated framework of $\py$ where it is harder to separate the many different effects. This initial study is useful to get some understanding of what to expect within a full $\py$ simulation. 

The simulated DGLAP evolution should occur at the same pace as the experimentally measured parton density scaling violations. Parton density parametrizations are fits to the $x$ evolution as a function of $\q$, so if the shower does not have the same rate of $x$ evolution the right parton density functions would not be obtained. A test of this is to pick partons according to the CTEQ5L parton density function at a low $\q_0$, evolve to a higher $\q$ using the toy model and then compare to an $x$-distribution generated according to the CTEQ5L parton density function at the higher $\q$. These $x$-distributions look close to identical, which is comforting. 

The next step is to do backwards evolution. To avoid the problem of quarks originating from gluons in the toy model, the backwards evolution is done only for valence quarks, where the same quark can be traced through the branchings back to the lower scale $\q_0$. The same test as for forward evolution can be done, this time comparing only the $x$-distributions for valence quarks and doing the evolution from the higher scale $\q$ to the lower $\q_0$. Again the result is comforting.

At this point the effect of angular ordering can be studied. The approximate opening angle of each emission is calculated using eq.~(\ref{eq:ang}) and an requirement, that the opening angle of an emission must be smaller than the previous one in the backward evolution, is introduced. This slows down the rate of emissions and consequently the rate of the evolution in $x$. To compensate for this $\lqcd$ can be raised. To quantify the increase in $\lqcd$ needed to compensate for angular ordering, $x$ distributions is generated by backwards evolution from an higher scale $\q = m_{\mrm{Z}}^2$ (to be able to compare to $\mrm{Z}^0$ production in $\py$) down to a lower scale $\q_0$. The backwards evolution is done with the center of mass energy set to 1800~Gev, as is the case for the Tevatron. At first the distribution is generated without the angular ordering requirement at the $\lqcd$ value 0.192~Gev used in the tuning of the CTEQ5L parton densities, then distributions are generated with the angular ordering requirement at different values of $\lqcd$. The mean values of $x$ can be plotted as a function of $\lqcd$ and at some value of $\lqcd$ the mean value of $x$ is the same as without angular ordering using $\lqcd$~=~0.192~Gev. According to the toy model algorithm, this new $\lqcd$ needed to compensate for angular ordering is 0.221~Gev. As this value is obtained looking only at the mean value of $x$, it is necessary to check that the entire $x$-spectra look alike. This is done in Fig.~\ref{fig:comp}. 

It should be noted that this study of the increase in $\lqcd$ needed to compensate for angular ordering in the toy model only includes valence quarks in the backwards evolution. After this initial studies in the toy model, more thorough studies of the effects of the corrections to DGLAP evolution on the $x$ evolution has been made using $\py$ version 6.220, but based on the same principles.

\subsubsection{$\py$ studies} 
\label{subsubsec-pst}

The same way of quantifying the increase in $\lqcd$ needed to compensate for angular ordering can be used, with $\py$ doing the backwards evolution. If this is done for $\mrm{Z}^0$ production at the Tevatron the plot in Fig.~\ref{fig:tevz} is obtained. According to $\py$ the $\lqcd$ needed is 0.224~Gev, not very far from the value given by the toy model. As in Fig.~\ref{fig:comp} the shape of parton densities also agree well.

This method can be used to find the $\lqcd$ values needed to compensate for all three corrections for two different processes, both at the Tevatron and LHC. The Tevatron is a p$\pbar$ collider with a center of mass energy of 1800~Gev and the LHC is a pp collider under construction with a planned center of mass energy of 14000~Gev. It is interesting to see how dependent the increase in $\lqcd$ is on the center of mass energy and on the generated process. The processes, q$\qbar \rightarrow$ $\mrm{Z}^0$ and gg~$\rightarrow$~$\mrm{H}^0$, are chosen to study the differences between having q$\qbar$ and gg in the initial state. $\mrm{H}^0$ denotes the Higgs boson, which is required within the Standard Model but has not yet been found. The Higgs mass is set to 120~Gev, which is somewhere in the neighbourhood of where it is expected to be. The results are presented in Table~\ref{tab:lambda}.

\begin{table}[h]
\begin{center}
\begin{tabular}{|l|c|c|c|c|} \hline                    
                   & \multicolumn{2}{|c|}{Tevatron}    & \multicolumn{2}{|c|}{LHC}    \\ \hline
                   & q$\qbar \rightarrow$ $\mrm{Z}^0$  & gg $\rightarrow$ $\mrm{H}^0$ & q$\qbar \rightarrow$ $\mrm{Z}^0$  & gg $\rightarrow$ $\mrm{H}^0$  \\ \hline
angular ordering   & 0.224 GeV   & 0.239 GeV & 0.221 GeV & 0.236 GeV \\ \hline
the condition that $\uhat < 0$   & 0.26 GeV   & 0.38 GeV & 0.21 GeV & 0.31 GeV \\ \hline
$\as$($\pt^2$) instead of $\as$($\q$)  & 0.28 GeV   & 0.30 GeV & 0.26 GeV & 0.27 GeV \\ \hline
sum of all effects  & 0.31 GeV   & 0.50 GeV & 0.29 GeV & 0.46 GeV \\ \hline
\end{tabular}
\caption{The value of $\lqcd$ needed to compensate for different corrections for q$\qbar \rightarrow$ $\mrm{Z}^0$ and gg~$\rightarrow$~$\mrm{H}^0$ at the Tevatron and LHC.}
\label{tab:lambda}
\end{center}
\end{table}

The increase to compensate for angular ordering appears to be almost independent of the center of mass energy and not very dependent of which process is generated. It would of course be convenient to be able to use the same $\lqcd$ value independently of the center of mass energy and particles in the initial state of the process studied. However for the other corrections the values seem to fluctuate more. 

$\lqcd$ needs to be increased more when the initial state is gg, compared to having the initial state q$\qbar$. This is not totally unexpected, since quark branchings and gluon branchings behave differently. The quark in the branching q~$\rightarrow$~qg tends to take a large momentum fraction. This, together with the ordering in $\q$, means that the constraints on the DGLAP evolution are often fulfilled. In the branching g~$\rightarrow$~gg one of the gluons tends to take a large momentum fraction and the other a small momentum fraction. If the gluon of the continued evolution takes a small momentum fraction, the constraints are not as naturally fulfilled as for the case with quarks. It would therefore be possible to associate different effective $\lqcd$ with different branching processes, but we did not pursue it here. There is also a difference in the increase in $\lqcd$ needed between the different values of center of mass energy. This difference is not so easy to explain, but considerably smaller.

\subsection{The effect of an increased $\lqcd$ value on $\pt$-spectra and primordial $\kt$} 
\label{subsec-pt}
\begin{figure}[t]
\begin{center}
\mbox{\epsfig{file=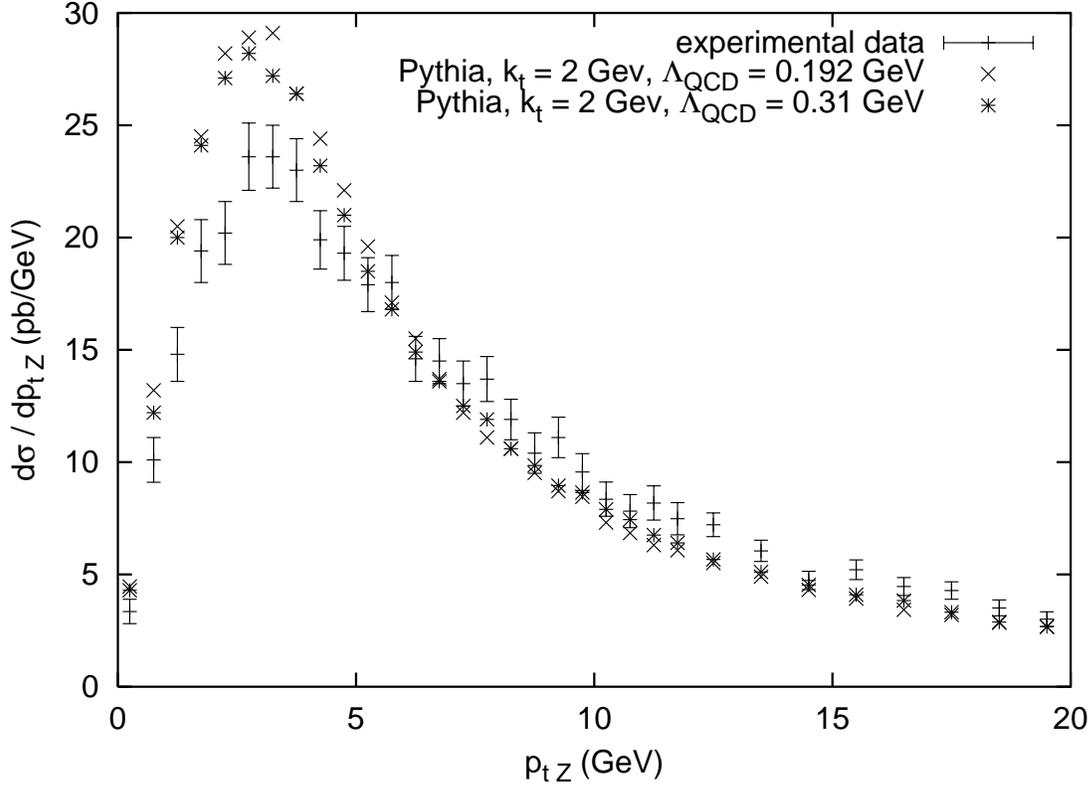,width=15cm}}
\caption{$p_{\perp \mrm{Z}}$-spectra for the process q$\qbar$~$\rightarrow$~$\mrm{Z}^0$ at the Tevatron, generated by $\py$ and experimental data. The $\py$ predictions are normalized to the total experimental cross section.}
\label{fig:exp}
\end{center}
\end{figure}

\begin{figure}[t]
\begin{center}
\mbox{\epsfig{file=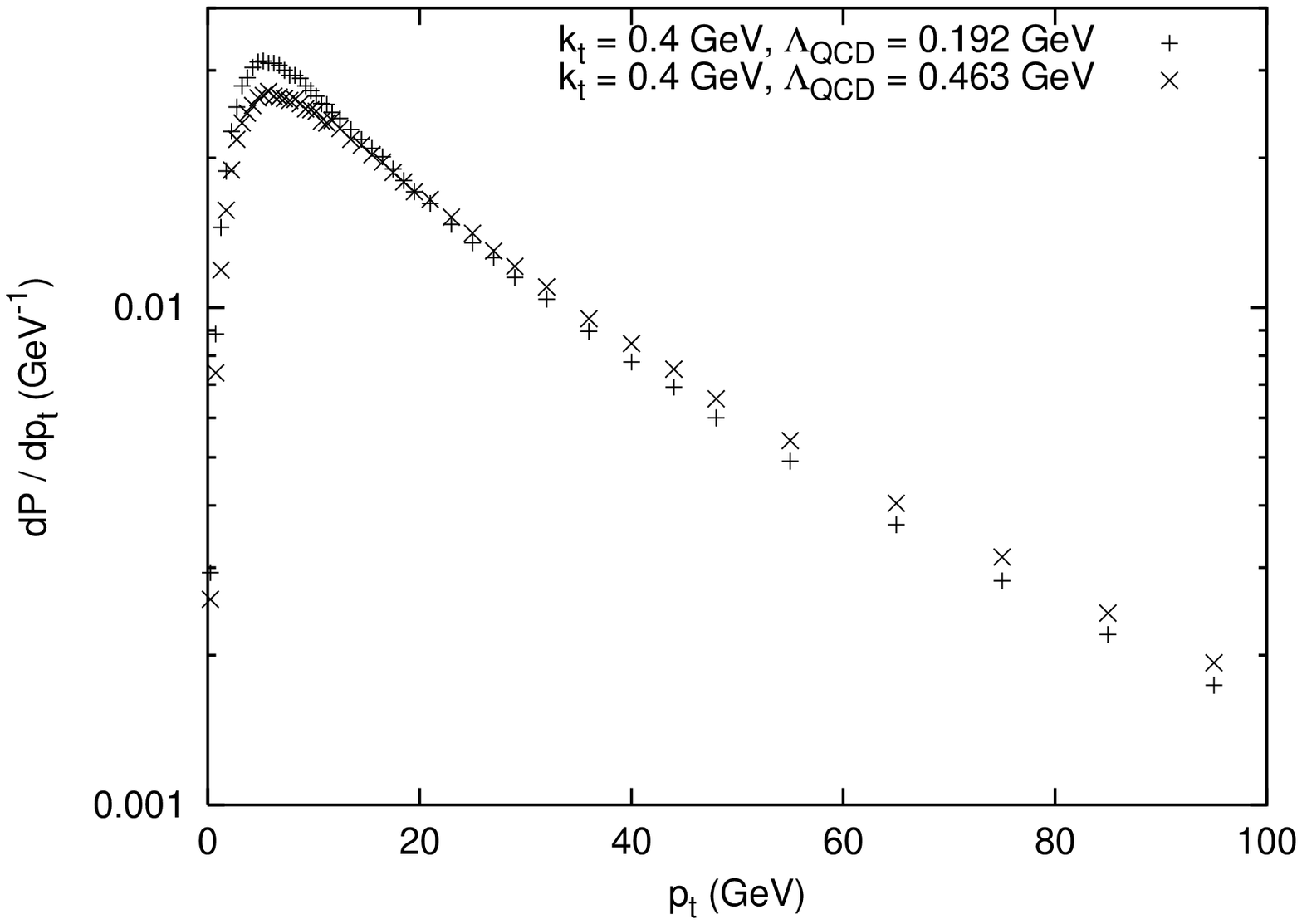,width=15cm}}
\caption{$p_{\perp \mrm{H}}$-spectra for the process gg~$\rightarrow$~$\mrm{H}^0$ at LHC generated by $\py$. The Higgs mass is assumed to be 120~Gev.}
\label{fig:lhch}
\end{center}
\end{figure}

In Fig.~\ref{fig:exp} $p_{\perp \mrm{Z}}$-spectra for the process q$\qbar$~$\rightarrow$~$\mrm{Z}^0$ at the Tevatron are shown. The $p_{\perp \mrm{Z}}$-spectra generated by Pythia, with $\lqcd$~=~0.192~Gev and $\lqcd$~=~0.31~Gev, are compared with experimental data \cite{Affolder:1999jh}. The increase in $\lqcd$ does not seem to affect the $p_{\perp \mrm{Z}}$-spectra much. To understand this it could be interesting to look at the effect of $\lqcd$ on $\as$ at a typical scale of the parton shower, $\q= 100 \,\mrm{GeV^2}$. When $\lqcd$ is increased from 0.192~Gev to 0.31 Gev, $\as(\q = 100 \,\mrm{GeV^2})$ is increased by approximately 10$\%$. The peak position of the $p_{\perp \mrm{Z}}$-spectra is shifted towards higher values by only approximately 1$\%$. This is because the $\pt$-kicks caused by the emissions are in different directions in the $xy$-plane (transverse to the direction of the beam), which means that they partly can cancel each other. If you instead look at the $E_{\perp}$ spectra, the sum of the absolute value of the $\pt$-kicks caused by emissions, the peak position is indeed shifted towards higher values by the expected 10$\%$.

Formally the $\pt$-kicks corresponds to emission of jets, so in theory the energy of these jets could be measured to get an experimental value for $E_{\perp}$. Unfortunately, this alternative measure of activity in $\mrm{Z}^0$ events is less accessible experimentally, since also other physics could contribute to activity in an event.

The absent effect of the increased $\lqcd$ value on the $p_{\perp \mrm{Z}}$-spectra means that something else is needed to solve the problem with the too high value of the primordial $\kt$ needed in $\py$ to fit experimental data. In Fig.~\ref{fig:exp} it is also obvious that in the $p_{\perp \mrm{Z}}$-spectra generated the event rates at the peaks are too high. In Subsection~\ref{subsec-time} another approach is described. 

For gg~$\rightarrow$~$\mrm{H}^0$ at LHC the change in $\lqcd$ has a bigger effect on the $p_{\perp \mrm{H}}$-spectra, see Fig.~\ref{fig:lhch}. However these spectra is not affected by primordial $\kt$ very much, which can be understood as follows. The center of mass energy at LHC is much greater, which means more emissions. At each branching the primordial $\kt$ is split between the two particles and not much is left at the production of the Higgs particle, cf. eq.~(\ref{eq:kt}).  

\subsection{The effect of final-state showers of emitted particles with timelike virtuality on $\pt$-spectra} 
\label{subsec-time}
\begin{figure}[t]
\begin{center}
\mbox{\epsfig{file=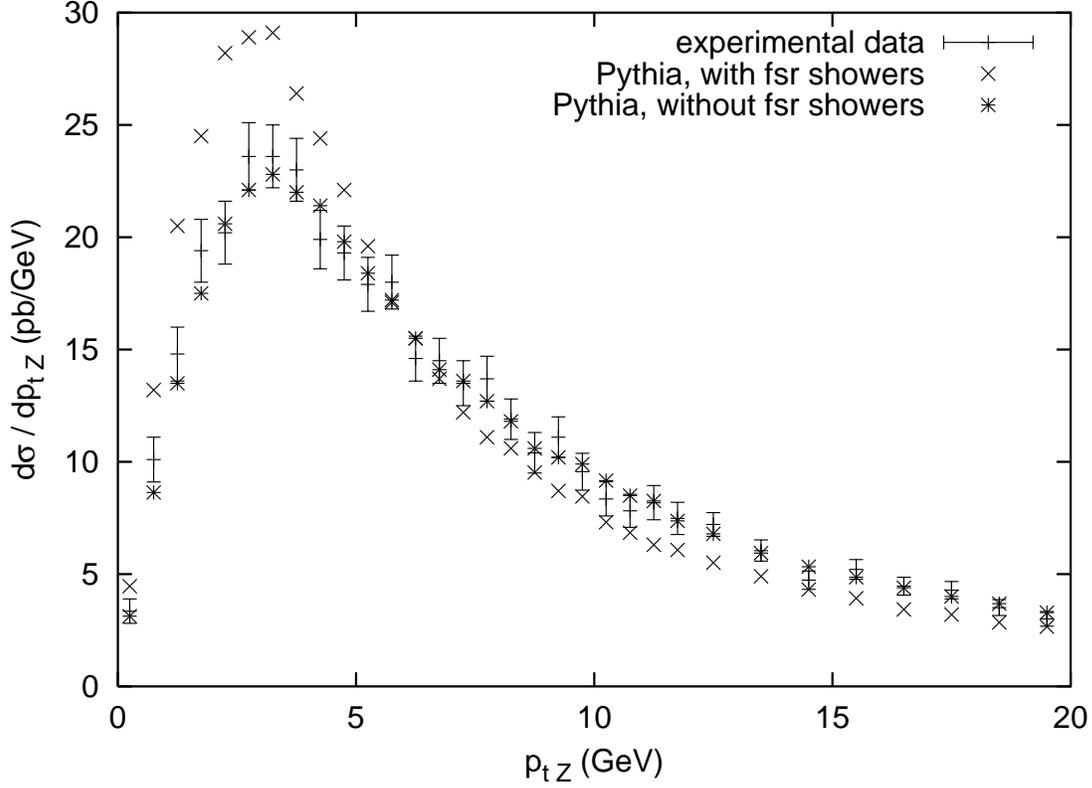,width=15cm}}
\caption{$p_{\perp \mrm{Z}}$-spectra for the process q$\qbar$~$\rightarrow$~$\mrm{Z}^0$ at the Tevatron. $\py$ generations with and without final-state showers of emitted particles with timelike virtuality, compared to experimental data. Both $\py$ generations use $\lqcd$ = 0.192~Gev and $\kt$ = 2~Gev.}
\label{fig:tid}
\end{center}
\end{figure}

As seen in Subsection~\ref{subsec-fin} final-state showers of emitted particles with timelike virtuality affect the $\pt$-spectrum of the produced particle. These timelike showers are supposed to have virtuality below the spacelike $\q$ at the emission. It could be argued that this means they occur too late to influence the spacelike parton of the branching. The recoil would instead be taken by colour-connected neighbour partons. Since both $\mrm{Z}^0$ and $\mrm{H}^0$ are colour neutral, their $\pt$-spectra would then be unaffected. If this is right, the $\pt$-spectra can be simulated by switching off final-state showers of emitted particles with timelike virtuality. A study of the recoiling jets would obviously require more, but we leave that aside here, since we do not need to study these jets for the purpose of this thesis. In Fig.~\ref{fig:tid} this has been done and the agreement with experimental data is much better. 

For gg~$\rightarrow$~$\mrm{H}^0$ at LHC the change in $\lqcd$ has an even bigger effect on the $\pt$-spectra, when final-state showers of emitted particles with timelike virtuality are switched off. The peak position is then shifted towards higher vaiues by approximately 50$\%$.

In a new shower algorithm \cite{new} the $\pt$-spectrum of the produced particle would not be affected by these timelike showers of emitted particles. This new algorithm is described in Subsection~\ref{subsec-new}. 

\subsection{$\pt$-spectra generated by the new algorithm} 
\label{subsec-new}
\begin{figure}[t]
\begin{center}
\mbox{\epsfig{file=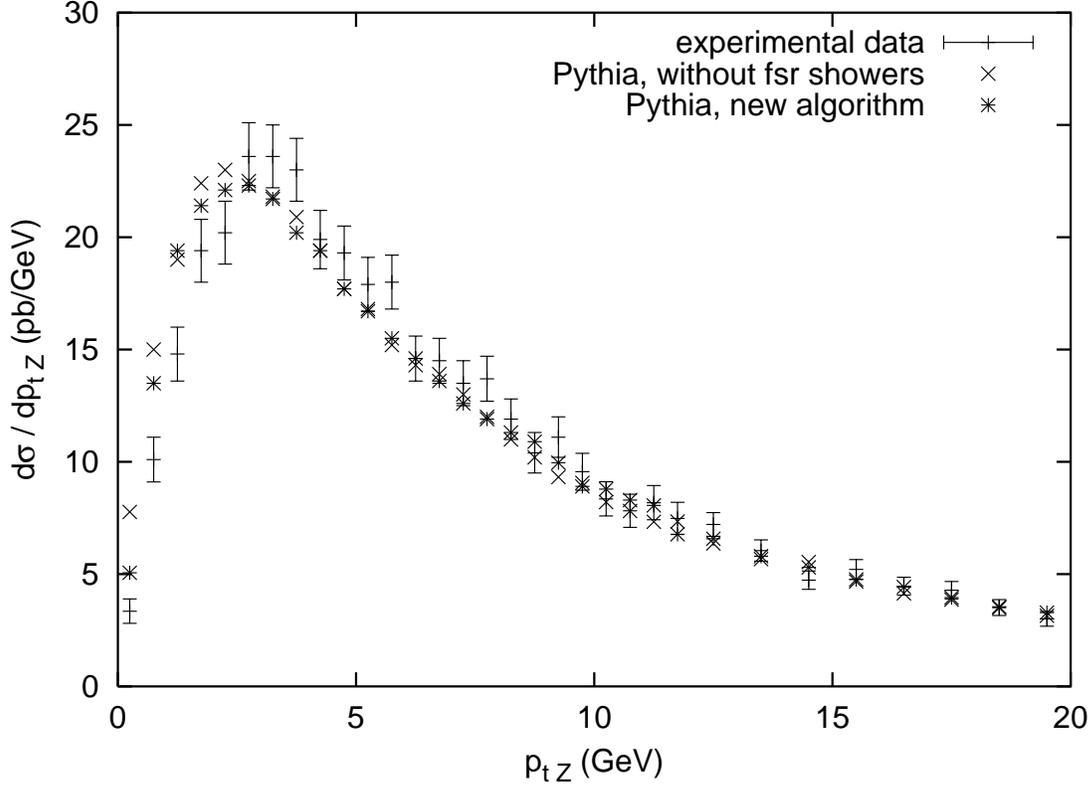,width=15cm}}
\caption{The $p_{\perp \mrm{Z}}$-spectrum generated by the new algorithm compared to the $p_{\perp \mrm{Z}}$-spectrum generated by $\py$ with final-state showers switched off. Both generations use $\lqcd$ = 0.192~Gev and $\kt$ = 0.4~Gev.}
\label{fig:ny}
\end{center}
\end{figure}

In the new algorithm the evolution is done in $\pt^2 = (1 - z) \q$ instead of $\q$. This is possible since

\begin{equation}
\frac{\dint \pt^2}{\pt^2} = \frac{(1 - z) \dint \q}{(1 - z) \q} = \frac{\dint \q}{\q}
\end{equation}
The ordering of emissions in $\pt^2$ gives a set of colour connected partons. These partons can then evolve a final state shower as a system. The $\mrm{Z}^0$-boson, as a colour neutral particle, is unaffected of colour dipoles.

To compare evolution in $\pt^2$ with evolution in $\q$, the $p_{\perp \mrm{Z}}$-spectrum generated by the new algorithm can be compared to the $p_{\perp \mrm{Z}}$-spectrum generated by $\py$ with final-state showers switched off. This is done in Fig.~\ref{fig:ny}. The two spectra look similar except for small $p_{\perp \mrm{Z}}$ where the new algorithm has better agreement with experimental data.

\begin{figure}
\begin{picture}(450,170)(0,0)
\SetWidth{2}
\Text(200,75)[]{\begin{tabular}{ccccccccccccc}
0.25 & &      &      & 78.6 & 68.9 & 79.7 & 81.4 & 78.5 &      &      &      &       \\
0.24 & &      &      & 81.3 & 64.6 & 68.8 & 70.1 & 70.5 &      &      &      &       \\
0.23 & & 77.2 & 81.1 & 74.0 & 69.2 & 64.6 & 64.5 & 66.1 & 64.7 & 61.0 & 76.4 & 90.1  \\
0.22 & & 83.6 & 80.5 & 63.4 & 66.1 & 68.0 & 61.0 & 62.8 & 50.0 & 64.5 & 63.3 & 77.2  \\
0.21 & & 78.3 & 72.7 & 71.0 & 69.9 & 60.3 & 60.0 & 53.7 & 57.9 & 60.9 & 55.9 & 71.8  \\
0.20 & & 78.9 & 85.5 & 74.6 & 59.0 & 57.7 & 57.9 & 48.7 & 51.8 & 45.2 & 53.6 & 55.4  \\
0.19 & & 97.4 & 83.6 & 80.3 & 64.2 & 55.2 & 56.5 & 49.7 & 45.8 & 46.1 & 40.7 & 55.6  \\
\multicolumn{13}{c}{}                                                                \\
     & & 0.2  & 0.4  & 0.6  & 0.8  & 1.0  & 1.2  & 1.4  & 1.6  & 1.8  & 2.0  & 2.2   \\
\end{tabular}}
\LongArrow(30,30)(450,30)
\LongArrow(30,30)(30,170)
\Text(15,160)[]{$\lqcd$}
\Text(425,15)[]{$\kt$}
\end{picture}
\caption{$\chi^2$ for fits to experimental data using the new algorithm with different values of $\lqcd$ and primordial $\kt$}
\label{fig:chi}
\end{figure}
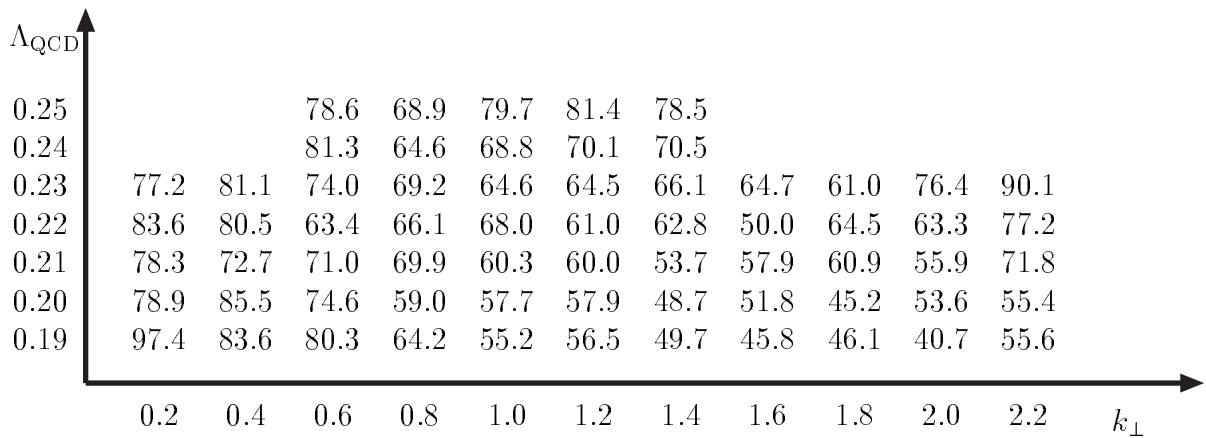

\begin{figure}[t]
\begin{center}
\mbox{\epsfig{file=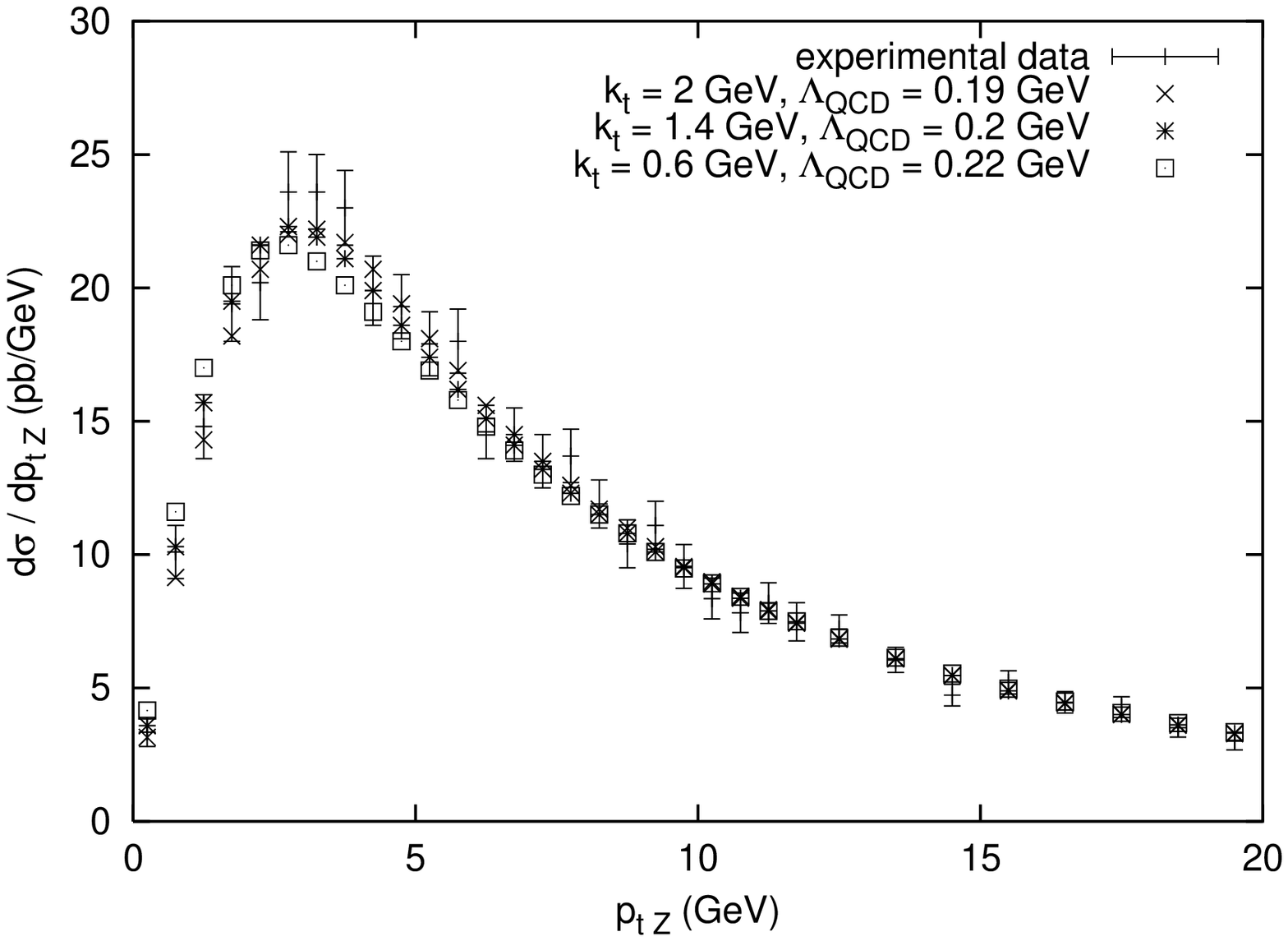,width=15cm}}
\caption{$p_{\perp \mrm{Z}}$-spectra generated by the new algorithm with $\lqcd$ and primordial $\kt$ values that give good $\chi^2$ values.}
\label{fig:chitwo}
\end{center}
\end{figure}

\begin{figure}[t]
\begin{center}
\mbox{\epsfig{file=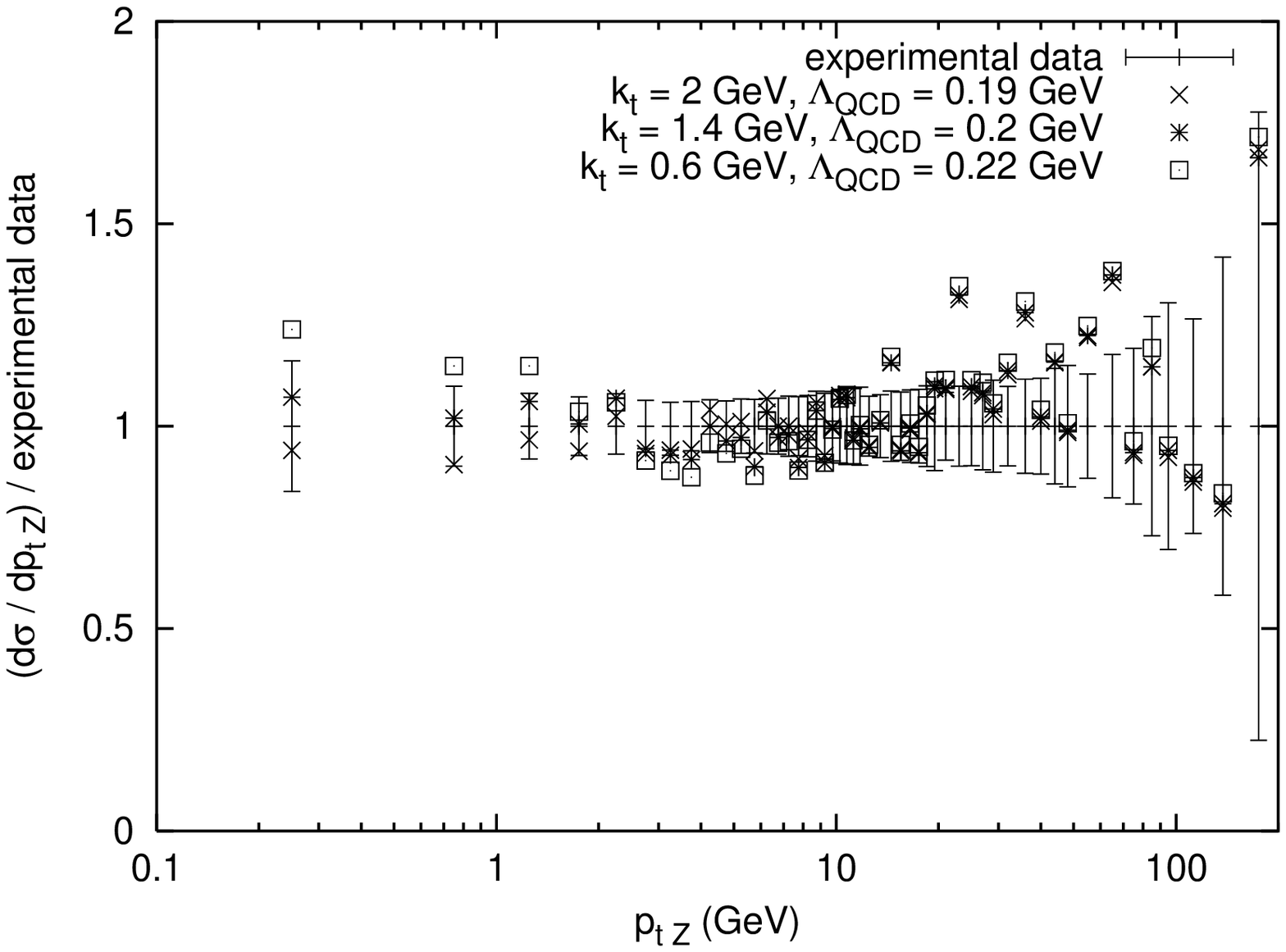,width=15cm}}
\caption{The same $p_{\perp \mrm{Z}}$-spectra as in Fig.~\ref{fig:chitwo}, divided by the experimental data points and over a wider range in $p_{\perp \mrm{Z}}$. The experimental errorbars are also divided by the experimental data points.}
\label{fig:chidiv}
\end{center}
\end{figure}

To find which values of $\lqcd$ and primordial $\kt$ that give the best fit to experimental data using the new algorithm, a $\chi^2$ test can be performed. $\chi^2$ is calculated by 

\begin{equation}
\chi^2 = \sum_{i=1}^n \frac{(\mrm{MC}_i - \mrm{data}_i)^2}{\sigma^2(\mrm{MC}_i) + \sigma^2(\mrm{data}_i)}
\end{equation}
where the summation is over the $n$ different experimental data points. In Fig.~\ref{fig:chi} $\chi^2$ values, generated with different values of $\lqcd$ and primordial $\kt$, are shown. The fit is best for low values of $\chi^2$. A perfect fit would give a $\chi^2$ value equal to the number of degrees of freedom (the number of data points minus the number of parameters that are to be determined), which in this case is 48. Due to fairly large statistical fluctuations and a lacking description of bin-to-bin correlations some of the values lie below this value. The $\chi^2$ values indicate that again the fit is best for $\lqcd \approx$~0.19~GeV and $\kt \approx$~2.0~GeV, but a trend can be seen that increasing $\lqcd$ gives better fits for lower values of $\kt$. 

In Fig.~\ref{fig:chitwo} $p_{\perp \mrm{Z}}$-spectra are generated with some $\lqcd$ and primordial $\kt$ values that give good $\chi^2$ values. These $p_{\perp \mrm{Z}}$-spectra are generated using 10\,000\,000 events, instead of the 100\,000 events used to generate the $\chi^2$ values in Fig.~\ref{fig:chi}. The $\chi^2$ values then become 46.5 for $\kt =$~2~Gev, $\lqcd =$~0.19~Gev, 51.4 for $\kt =$~1.4~Gev, $\lqcd =$~0.2~Gev and 74.6 for $\kt =$~0.6~Gev, $\lqcd =$~0.22~Gev. In Fig.~\ref{fig:chitwo} it can also be seen that the two $p_{\perp \mrm{Z}}$-spectra generated with $\kt =$~2~Gev, $\lqcd =$~0.19~Gev and $\kt =$~1.4~Gev, $\lqcd =$~0.2~Gev give good fits to the experimental data, while the one generated with $\kt =$~0.6~Gev, $\lqcd =$~0.22~Gev have too many events at low $p_{\perp \mrm{Z}}$ and too few around the peak.

Fig.~\ref{fig:chidiv} shows the same $p_{\perp \mrm{Z}}$-spectra as in Fig.~\ref{fig:chitwo}, divided by the experimental data points, over a wider range in $p_{\perp \mrm{Z}}$. The $p_{\perp \mrm{Z}}$-spectra seem to differ mainly for small values of $p_{\perp \mrm{Z}}$, which is what is expected when different values of primordial $\kt$ are used. In Fig.~\ref{fig:chidiv}, there is a trend that the $p_{\perp \mrm{Z}}$-spectra with lower values of primordial $\kt$ have more events with low $p_{\perp \mrm{Z}}$. A small effect of the change in $\lqcd$ can also be seen for high values of $p_{\perp \mrm{Z}}$ in Fig.~\ref{fig:chidiv}, where higher values of $\lqcd$ mean fewer events around the peak and more a bit above that.

At $p_{\perp \mrm{Z}}$ values a bit above the peak many of the generated points lie above the experimental data points. The fact that the points of the three different generations appear to fluctuate in unison owes to the point-to-point fluctuations in the data, where the statistics is much smaller than in the Monte Carlo studies. Again the two $p_{\perp \mrm{Z}}$-spectra generated with $\kt =$~2~Gev, $\lqcd =$~0.19~Gev and $\kt =$~1.4~Gev, $\lqcd =$~0.2~Gev fit the experimental data equally well, while the one with $\kt =$~0.6~Gev, $\lqcd =$~0.22~Gev have too many events with low $p_{\perp \mrm{Z}}$. However it should be noted that the points where the fit of the $p_{\perp \mrm{Z}}$-spectrum with $\kt =$~0.6~Gev, $\lqcd =$~0.22~Gev is considerably worse than the others are mainly the three first. A systematic error in the experimental data could affect closely lying points in the same way and thus bias the conclusions. 

The new algorithm does not yet include heavy quarks (b, c). When $\pt$-spectra generated by the ordinary $\py$ algorithm are studied, the exclusion of heavy quarks does not have any measurable effect, however. 
\section{Inclusion of heavy quarks in the new algorithm}
\label{sec-hq}

This section is an attempt to find a way of introducing heavy quarks in the new algorithm. No obvious way to do this is found and the work is under development. This is only a beginning of these studies. 

\subsection{q~$\rightarrow$~qg} 
\label{subsec-qqg}
\begin{figure}[t]
\begin{picture}(200,100)(0,0)
\SetWidth{2}
\ArrowLine(0,80)(60,40)
\ArrowLine(0,0)(60,40)
\ArrowLine(140,40)(200,80)
\ArrowLine(140,40)(200,0)
\DashLine(60,40)(140,40){5}
\Text(30,70)[l]{q}
\Text(100,50)[]{$\mrm{H}^0$}
\Text(30,10)[l]{$\qbar$}
\Text(160,70)[l]{$\eplus$}
\Text(160,10)[l]{$\eminus$}
\Text(0,90)[l]{a)}
\end{picture}
\begin{picture}(200,100)(-20,0)
\SetWidth{2}
\ArrowLine(0,80)(60,40)
\ArrowLine(0,0)(60,40)
\ArrowLine(140,40)(200,80)
\ArrowLine(140,40)(200,0)
\Gluon(30,60)(60,80){3}{3}
\DashLine(60,40)(140,40){5}
\Text(15,75)[l]{q}
\Text(100,50)[]{$\mrm{H}^0$}
\Text(30,10)[l]{$\qbar$}
\Text(160,70)[l]{$\eplus$}
\Text(160,10)[l]{$\eminus$}
\Text(0,90)[l]{b)}
\end{picture}
\caption{a) and b) q$\qbar$~$\rightarrow$~$\mrm{H}^0$~$\rightarrow$~$\eplus\eminus$ with and without the emission of a gluon. $\mrm{H}^0$ is a made-up scalar particle that has the couplings necessary for the processes.}
\label{fig:qqg}
\end{figure}
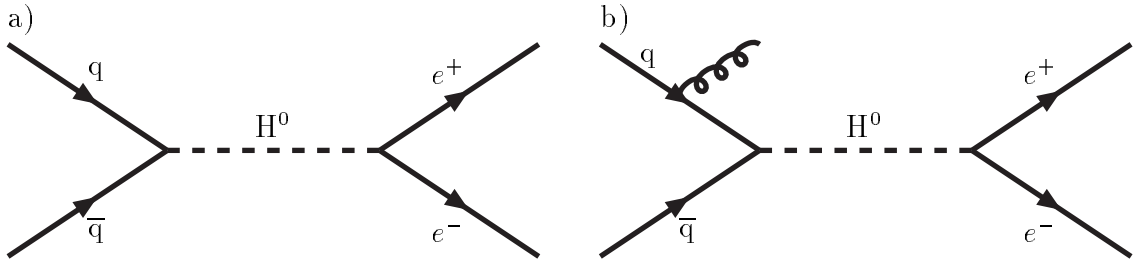

To derive the probability for a quark to emit a gluon, the cross section for a process that involves a quark emitting a gluon can be divided by the cross section for the same process without the emission of the gluon. The process used in this derivation, shown in Fig.~\ref{fig:qqg}, involves a made-up Higgs particle that couples to the particles needed to make the processes as simple as possible with a vertex factor of 1. This is allowed since the aim is to derive the probability of the quark emitting a gluon in the colinear limit where it does not depend on the rest of the process.

The needed matrix elements squared are calculated using CompHep \cite{Pukhov:1999gg}. One problem of technical character, is that CompHep require the Higgs to be unstable, so we have to generate an isotropic Higgs two-body decay. This makes the calculations more complicated. 

%
 
\subsubsection{Derivation of the DGLAP splitting function for massless quarks} 
\label{subsubsec-derqqg}
The cross section for the process in Fig.~\ref{fig:qqg}b is given by

\begin{equation}
\sigma = \int \dint x_1 \dint x_2 f(x_1) f(x_2) \, \hat{\sigma}_3 \, \dint \PS_3
\label{eq:sig}
\end{equation}
where $x_1$, $x_2$ are the fractions of the total proton momentum carried by the initial quarks, $f$ is the structure function of the proton (neglecting the $\q$ dependence), $\dint \PS_3$ is the integration over the phase space for the three final particles and the constituent cross section is given by

\begin{equation}
\hat{\sigma}_3 \propto \frac{\overline{|M_{2 \rightarrow 3}|^2}}{2 \shat} \delta (m_{\mrm{i}}^2 - m_{\mrm{H}}^2)
\label{eq:const}
\end{equation}
where $\overline{|M_{2 \rightarrow 3}|^2}$ is the matrix element squared and $m_{\mrm{i}}$ is the mass of the intermediate particle. 

\begin{figure}[h]
\begin{center}
\begin{picture}(200,100)(0,0)
\Oval(30,50)(20,10)(0)
\ArrowLine(37,40)(200,0)
\DashLine(37,60)(120,80){5}
\ArrowLine(120,80)(200,100)
\ArrowLine(120,80)(200,60)
\Text(190,90)[l]{1}
\Text(190,70)[l]{2}
\Text(140,25)[l]{3}
\Text(30,80)[l]{0}
\Text(80,60)[l]{i}
\end{picture}
\caption{The final particles of the process in Fig.~\ref{fig:qqg}b. 3 is the gluon, i is the intermediate made up Higgs boson, 1 and 2 are the $\eplus\eminus$-pair and 0 denotes the available energy or momenta from the collision.}
\label{fig:PS}
\end{center}
\end{figure}
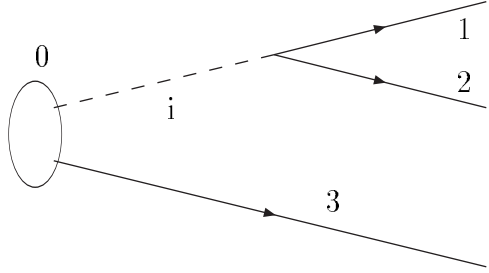

With the particles denoted according to Fig.~\ref{fig:PS}, the phase space integral is given by

\begin{equation}
\dint \PS_3 = \frac{1}{(2 \pi)^9} \frac{\dint^3 \overline{p}_1}{2 E_1} \frac{\dint^3 \overline{p}_2}{2 E_2} \frac{\dint^3 \overline{p}_3}{2 E_3} \,  (2 \pi)^4 \, \delta^{(4)} (P_0 - P_1 - P_2 - P_3)
\end{equation}  
where $P$ denotes four-momentum. This can be rewritten as

\begin{equation}
\dint \PS_3 = \dint \PS_2 \cdot \frac{1}{4 \cdot (2 \pi)^2} \int \dint m_{\mrm{i}}^2 \, \frac{\dint \that}{x_1 x_2 s}
\label{eq:phfin}
\end{equation}
where $\dint \PS_2$ is the phase space integral for the process in Fig.~\ref{fig:qqg}a and $\that$ is the Mandelstam variable for the process q$\qbar$~$\rightarrow$~$\mrm{H}^0$g. The cross section given in eq.~(\ref{eq:sig}) then becomes 

\begin{equation}
\sigma \propto \frac{1}{4 \cdot (2 \pi)^2} \int \dint x_1 \dint x_2 f(x_1) f(x_2) \frac{\overline{|M_{2 \rightarrow 3}|^2}}{2 \shat} \, \dint m_{\mrm{i}}^2 \, \frac{\dint \that}{x_1 x_2 s} \, \dint \PS_2 \, \delta (m_{\mrm{i}}^2 - m_{\mrm{H}}^2)
\label{eq:sigm}
\end{equation}
 
In the parton shower approximation, the cross section for the process in Fig.~\ref{fig:qqg}b can also be calculated using the cross section for the process in Fig.~\ref{fig:qqg}a and the DGLAP splitting function for q~$\rightarrow$~qg

\begin{equation}
\sigma \propto \int \dint x_1 \dint x_2 f(x_1) f(x_2) \frac{\overline{|M_{2 \rightarrow 2}|^2}}{2 \shat} \, \dint z \, \frac{\dint \q}{\q} \frac{\as}{2 \pi} P_{\mrm{q \rightarrow qg}}(z)  \, \dint \PS_2 \, \delta (x_1 x_2 z s - m_{\mrm{H}}^2)
\label{eq:sigd}
\end{equation}
In the collinear limit where 
 
\begin{equation}
m_{\mrm{i}}^2 = x_1 x_2 z s \Rightarrow \dint m_{\mrm{i}}^2 = x_1 x_2 s \dint z
\end{equation}
eq.~(\ref{eq:sigm}) becomes

\begin{equation}
\sigma \propto \frac{1}{4 \cdot (2 \pi)^2} \int \dint x_1 \dint x_2 f(x_1) f(x_2) \frac{\overline{|M_{2 \rightarrow 3}|^2}}{2 \shat} \, \dint z \, \dint \that \, \dint \PS_2 \, \delta (x_1 x_2 z s - m_{\mrm{H}}^2)
\label{eq:sigme}
\end{equation} 
The condition that the two expressions for the cross section, eq.~(\ref{eq:sigd}) and eq.~(\ref{eq:sigme}), should be equal gives, after some further considering

\begin{equation}
P_{\mrm{q \rightarrow qg}}(z) = \frac{\q}{8 \pi \as} \frac{\overline{|M_{2 \rightarrow 3}|^2}}{\overline{|M_{2 \rightarrow 2}|^2}}
\end{equation}
The matrix elements squared can be calculated, e.g. using CompHep. The process in Fig.~\ref{fig:qqg}b can not be separated from the process where the other quark emits the gluon. The matrix elements must be added and then squared. In the collinear limit where the gluon is emitted in the direction of the first quark the matrix element squared of the second process disappears. The interference term however can contribute to the derived expression for the splitting function.  Using the CompHep matrix elements squared gives the following expression for the splitting function

\begin{equation}
P_{\mrm{q \rightarrow qg}}(z) = \frac{4}{3} \frac{\shat^2 (z^2 + 1) - Q^4}{\shat \, (\shat(1 - z) + Q^2)}
\label{eq:spf}
\end{equation}
In the collinear limit, where $\q$ is small compared to $\shat$, eq.~(\ref{eq:spf}) goes to the DGLAP splitting function in eq.~(\ref{eq:spl}).

\subsubsection{Massive quarks} 
\label{subsubsec-massqqg}

The corresponding calculation can be done for the case when the quark emitting the gluon in the process in Fig.\ref{fig:qqg}b has a non-negligible mass. For a c quark the calculation gives

\begin{eqnarray}
P_{\mrm{c \rightarrow cg}}(z) &=& \frac{4\q}{3} \frac{(3 z^2 m_{\mrm{c}}^2 + z^2 \q - 2 z m_{\mrm{c}}^2 + m_{\mrm{c}}^2 + \q) \shat -2 m_{\mrm{c}}^2 \q - 2 m_{\mrm{c}}^4 z}{(m_{\mrm{c}}^2 + Q^2)^2 \, ((1 - z) \shat - m_{\mrm{c}}^2 - Q^2)} \nonumber \\
\nonumber \\
\Rightarrow \frac{\dint \mathcal{P}_{\mrm{c \rightarrow cg}}}{\dint \q} &=& \frac{\as}{2 \pi} \frac{4}{3} \frac{(3 z^2 m_{\mrm{c}}^2 + z^2 \q - 2 z m_{\mrm{c}}^2 + m_{\mrm{c}}^2 + \q) \shat -2 m_{\mrm{c}}^2 \q - 2 m_{\mrm{c}}^4 z}{(m_{\mrm{c}}^2 + Q^2)^2 \, ((1 - z) \shat - m_{\mrm{c}}^2 - Q^2)}
\label{eq:spfc}
\end{eqnarray}
Since the new algorithm is ordered in $\pt^2$, what is needed is really $\frac{\dint \mathcal{P}_{\mrm{c \rightarrow cg}}}{\dint \pt^2}$ instead of $\frac{\dint \mathcal{P}_{\mrm{c \rightarrow cg}}}{\dint \q}$. For the case of a c quark emitting a gluon $\pt^2$ as a function of $\q$ becomes

\begin{eqnarray}
\pt^2 &=& \frac{((\shat - m_{\mrm{c}}^2)(1 - z) - \q)\q \shat}{(\shat - m_{\mrm{c}}^2)^2} \nonumber \\
&\Rightarrow& \q = \frac{(\shat - m_{\mrm{c}}^2) (1 - z)}{2} - (\shat - m_{\mrm{c}}^2) \sqrt{\frac{1}{4} (1 - z)^2 - \frac{\pt^2}{\shat}} \nonumber \\
&\Rightarrow& \frac{\dint \q}{\dint \pt^2} = - \frac{1}{2} \frac{\shat - m_{\mrm{c}}^2}{\sqrt{\frac{1}{4} (1 - z)^2 - \frac{\pt^2}{\shat}}} 
\end{eqnarray}
Neglecting the $Q^4$ term gives

\begin{eqnarray}
\q &=& \frac{(\shat - m_{\mrm{c}}^2) \pt^2}{(1 - z) \shat} \nonumber \\
&\Rightarrow& \frac{\dint \q}{\dint \pt^2} = \frac{\shat - m_{\mrm{c}}^2}{(1 - z) \shat} 
\end{eqnarray}
Even with these simplified expressions for $\q$ and $\frac{\dint \q}{\dint \pt^2}$, the expression for $\frac{\dint \mathcal{P}_{\mrm{c \rightarrow cg}}}{\dint \pt^2}$ gets very complicated. 

Complexity in itself is not an absolute hindrance, however: if a simple upper estimate can be found, standard rejection techniques can be used to obtain results in agreement with the full expression. It therefore remains to find some suitable such estimate in a $\pt$-oriented framework.
\subsection{g~$\rightarrow$~q\boldmath$\qbar$} 
\label{subsec-gqqbar}

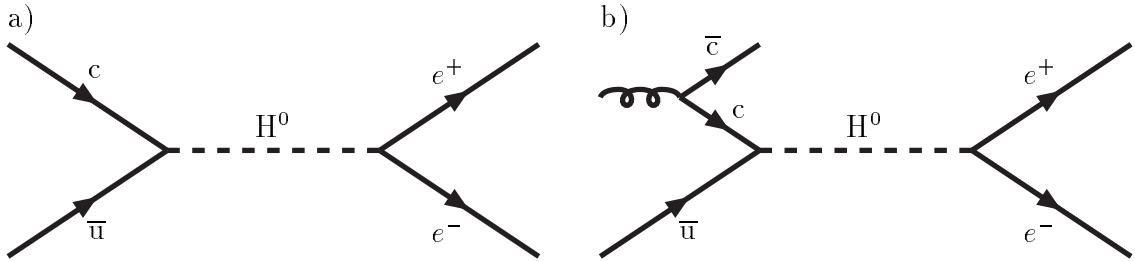
\begin{figure}[t]
\begin{picture}(200,100)(0,0)
\SetWidth{2}
\ArrowLine(0,80)(60,40)
\ArrowLine(0,0)(60,40)
\ArrowLine(140,40)(200,80)
\ArrowLine(140,40)(200,0)
\DashLine(60,40)(140,40){5}
\Text(30,70)[l]{c}
\Text(100,50)[]{$\mrm{H}^0$}
\Text(30,10)[l]{$\ubar$}
\Text(160,70)[l]{$\eplus$}
\Text(160,10)[l]{$\eminus$}
\Text(0,90)[l]{a)}
\end{picture}
\begin{picture}(200,100)(-20,0)
\SetWidth{2}
\ArrowLine(30,60)(60,40)
\ArrowLine(30,60)(60,80)
\ArrowLine(0,0)(60,40)
\ArrowLine(140,40)(200,80)
\ArrowLine(140,40)(200,0)
\Gluon(0,60)(30,60){3}{2}
\DashLine(60,40)(140,40){5}
\Text(50,55)[l]{c}
\Text(100,50)[]{$\mrm{H}^0$}
\Text(40,80)[l]{$\cbar$}
\Text(30,10)[l]{$\ubar$}
\Text(160,70)[l]{$\eplus$}
\Text(160,10)[l]{$\eminus$}
\Text(0,90)[l]{b)}
\end{picture}
\caption{q$\qbar$~$\rightarrow$~$\mrm{H}^0$~$\rightarrow$~$\eplus\eminus$. In b) the quark comes from a gluon splitting into a q$\qbar$-pair. $\mrm{H}^0$ is again a made-up scalar particle that has the coupling necessary for the processes.}
\label{fig:gqqbar}
\end{figure}    

To derive the probability for a gluon to split into a c$\cbar$ pair, the same method as in Subsection~\ref{subsec-qqg} can be used, this time with the processes shown in Fig.~\ref{fig:gqqbar}. The following expression is obtained when, again using CompHep, the matrix element squared for the process in Fig.~\ref{fig:gqqbar}b is divided by the matrix element squared for the process in Fig.~\ref{fig:gqqbar}a

\begin{eqnarray}
4 \pi \as \frac{(2 z^2 m_{\mrm{c}}^2 + 2 z^2 \q -2 z \q + m_{\mrm{c}}^2 + \q) \shat}{(m_{\mrm{c}}^2 + \q)^2 \, (\shat - m_{\mrm{c}}^2)} &+& 4 \pi \as \frac{2 z Q^4 - 2 m_{\mrm{c}}^2 \q - 2 Q^4 - 2 m_{\mrm{c}}^4 - 2 m_{\mrm{c}}^4 z}{(m_{\mrm{c}}^2 + \q)^2 \, (\shat - m_{\mrm{c}}^2)} + \nonumber \\
\nonumber \\
+ \,\,\, 4 \pi \as \frac{m_{\mrm{c}}^6 + m_{\mrm{c}}^4 \q + m_{\mrm{c}}^2 Q^4 + Q^6}{\shat (m_{\mrm{c}}^2 + \q)^2 \, (\shat - m_{\mrm{c}}^2)}&&
\label{eq:gqq}
\end{eqnarray}
If the fact that there is a massive c quark in the initial state of the process in Fig.~\ref{fig:gqqbar}a is taken into account the denominator in Eq.~\ref{eq:const} should be $2(\shat - m_{\mrm{c}}^2)$ instead of $2\shat$. This means that the expression in Eq.~\ref{eq:gqq} should be multiplied with $\frac{(\shat - m_{\mrm{c}}^2)}{\shat}$, which gives the following expression

\begin{eqnarray}
4 \pi \as \frac{\q(z^2 + (1 - z)^2) + (1 + 2z^2) m_{\mrm{c}}^2}{(m_{\mrm{c}}^2 + \q)^2} &-& \nonumber \\
\nonumber \\
- \, 8 \pi \as \frac{(1 - z)Q^4 + m_{\mrm{c}}^2 \, (m_{\mrm{c}}^2 + \q) + m_{\mrm{c}}^4 z}{\shat (m_{\mrm{c}}^2 + \q)^2} &+& ...
\label{eq:gcc}
\end{eqnarray}
where the first term takes the form of the DGLAP splitting function when $m_{\mrm{c}}^2$ is set to 0.

In the process in Fig.~\ref{fig:gqqbar}b there is a massive $\cbar$ quark in the final state. This means that the phase space integral gets more complicated

\begin{equation}
\dint \PS_3 = \dint \PS_2 \cdot \frac{1}{4 \cdot (2 \pi)^2} \int \frac{\sqrt{(x_1 x_2 s - m_i^2 + m_{\mrm{c}}^2)^2 - 4 x_1x_2 s m_{\mrm{c}}^2}}{x_1 x_2 s - m_i^2 + m_{\mrm{c}}^2} \, \dint m_{\mrm{i}}^2 \, \frac{\dint \that}{x_1 x_2 s}
\label{eq:phcomp}
\end{equation}

Again it remains to find a simple upper estimate as a starting point for a more precise inclusion of mass effects.

\section{Summary}
\label{sec-sum}

To explore perturbative and nonperturbative effects in transverse momentum generation $\pt$-spectra have been studied. More specificaly, the influence of corrections in $\py$ to pure leading log DGLAP evolution on $x$ evolution and generation of $\pt$-spectra has been investigated. This has been done in hope that the reason for the need of a primordial $\kt$ much larger than expected might be found. 

To compensate for the corrections to pure leading log DGLAP evolution in $\py$, $\lqcd$ needs to be raised to the values given in Table~\ref{tab:lambda}. For q$\qbar$~$\rightarrow$~$\mrm{Z}^0$ at the Tevatron this increase in $\lqcd$ does not affect the generated $p_{\perp \mrm{Z}}$-spectra very much. This can be understood as the effect of $\pt$ kicks of the branchings in different directions in the plane transverse to the beam direction partly canceling each other. The generated $p_{\perp \mrm{Z}}$-spectra do not fit the experimental data very well. For gg~$\rightarrow$~$\mrm{H}^0$ at LHC a changed $\lqcd$ has a larger effect on the generated $p_{\perp \mrm{H}}$-spectra, due to higher center of mass energy and consequently more branchings. In this case the primordial $\kt$ of the initial partons does not affect the generated $p_{\perp \mrm{H}}$-spectra very much, since it is split between the partons of the many branchings.

Final-state showers of emitted particles with timelike virtuality affect the generated $\pt$-spectra. It could be argued that these showers occur too late to influence the spacelike parton of the branching, since they are supposed to have virtuality below the virtuality of the branching. When these are turned off the fit to the experimental data for q$\qbar$~$\rightarrow$~$\mrm{Z}^0$ at the Tevatron is much better.

In a new shower algorithm the branchings are ordered in $\pt^2$ instead of $\q$. This gives a set of colour connected partons which can then undergo final state showers as a system. These final state showers would not affect the produced $\mrm{Z}^0$, since it is a colour neutral particle. In this algorithm the increase in $\lqcd$ has a larger effect on the generated $p_{\perp \mrm{Z}}$-spectra. When three different $p_{\perp \mrm{Z}}$-spectra, using values of $\lqcd$ and primordial $\kt$ that give good $\chi^2$ values, are studied the fit is equally good for the two spectra using $\kt =$~2~Gev, $\lqcd =$~0.19~Gev and $\kt =$~1.4~Gev, $\lqcd =$~0.2~Gev respectively, while the fit is a bit, but not very much, worse for the one using $\kt =$~0.6~Gev, $\lqcd =$~0.22~Gev. The problem with a large value of primordial $\kt$ is not completely solved but reduced.

The studies to include heavy quarks in the new algorithm are under development. No obvious way of doing this has yet been found.

\pagebreak

\end{document}